\documentclass[letterpaper]{article}
\usepackage[T1]{fontenc}
\usepackage{geometry}
\usepackage{bm}
\usepackage{setspace}
\usepackage{mathtools}
\usepackage[sorting=none]{biblatex}
\usepackage{acro}

\DeclareAcronym{cpu}{
  short = CPU ,
  long  = Central Processing Unit
}

\DeclareAcronym{gpu}{
  short = GPU ,
  long  = Graphics Processing Unit
}

\DeclareAcronym{maze}{
    short = MaZe,
    long = Mass-Zero constrained dynamics
}

\DeclareAcronym{pmaze}{
    short = Poisson MaZe,
    long = Poisson Mass-Zero constrained dynamics,
    first-style = short
}

\DeclareAcronym{p3maze}{
    short = P3MaZe,
    long = Particle-particle particle-mesh MaZe,
    first-style = short
}

\DeclareAcronym{fft}{
    short = FFT,
    long = Fast Fourier transform
}

\DeclareAcronym{p3m}{
    short = P3M,
    long = Particle-Particle Particle-Mesh
}

\DeclareAcronym{md}{
    short = MD,
    long = molecular dynamics
}

\DeclareAcronym{pbc}{
short=PBC,
long= periodic boundary conditions
}

\DeclareAcronym{vacf}{
short=VACF,
long=velocity autocorrelation function
}

\DeclareAcronym{cacf}{
short=CACF,
long=current autocorrelation function
}

\DeclareAcronym{spc}{
short=SPC/Fw,
long=simple point-charge flexible water
}
\usepackage[hidelinks]{hyperref}
\usepackage{graphicx}
\usepackage{float}
\newfloat{scheme}{htbp}{los}
\floatname{scheme}{Scheme}
\floatname{chart}{Chart}
\newfloat{graph}{htbp}{loh}

\usepackage{chemformula} % Formulas using \ch{}
\usepackage[version = 4]{mhchem} % Formulas using \ce{}
\usepackage{color}

\usepackage{xcolor}
\usepackage{amssymb}
\usepackage{authblk}
\author[1]{Federica Troni}
\author[1]{Violette Gontran}
\author[1]{Davide Grassano}
\author[1,2]{Sara Bonella*}
\affil[1]{Centre Européen de Calcul Atomique et Moléculaire (CECAM), Ecole Polytechnique Fédérale de Lausanne, 1015 Lausanne, Switzerland}
\affil[2]{National Centre for Computational Design and Discovery of Novel Materials (MARVEL), Ecole Polytechnique Fédérale de Lausanne,CH-1015 Lausanne, Switzerland}

\title{P3MaZe: a Mass-Zero constrained-dynamics formulation of particle--mesh electrostatics}
\date{*Email: sara.bonella@epfl.ch}
\begin{document}

\maketitle

\begin{abstract}
We introduce \ac{p3maze}, a real-space particle--mesh electrostatic method that combines the standard short-range/long-range decomposition of \ac{p3m} electrostatics with the \ac{maze} framework. In this formulation, the smooth long-range electrostatic potential is represented on a mesh as a zero-inertia auxiliary field, while the discretized Poisson equation is enforced as a holonomic constraint during molecular dynamics. By retaining the standard \ac{p3m} decomposition, \ac{p3maze} preserves the systematic accuracy controls associated with the real-space cutoff, the Ewald splitting, the mesh spacing, and the charge assignment procedure, while replacing the conventional multigrid Poisson solver by a constrained correction problem. The method is validated for molten NaCl and \ac{spc}. Values of the potential, structural, translational, collective, and rotational dynamical observables are in quantitative agreement with those obtained with established electrostatic methods, including real-space \ac{p3m}, and Ewald summation. We further show that, for the linear, time-independent Poisson operator considered here, the constrained correction problem is algebraically equivalent to a direct Poisson solve initialized by a chronological extrapolation of the mesh potential -- a connection not previously established for MaZe-based approaches. This result identifies the precise mechanism through which acceleration of the iterative long-range solve is achieved, and we exploit it systematically through higher-order predictors to further improve convergence, while retaining the expected linear scaling with system size. Because the underlying extrapolation is independent of the constrained-dynamics formalism, it can, of course, also be used directly to accelerate conventional iterative real-space Poisson solvers. 

\end{abstract}

\section*{Keywords}
Real--space particle--mesh electrostatics; Constrained dynamics; Molecular dynamics.

\section{Introduction}
\label{sec:introduction}
%%%%%%%%%%%%%

The evaluation of long-range electrostatic interactions remains one of the main challenges in large-scale classical \ac{md} simulations. Unlike short-ranged van der Waals interactions, these interactions cannot be treated by a simple local truncation in a controlled manner~\cite{frenkelsmit2023}. Under \ac{pbc}, the lattice sum is conditionally convergent, and the electrostatic response contains both a singular short-range component, dominated by local pairwise interactions, and a long-range component that remains collective over the periodic system~\cite{Allen-Tildesley}. Highly efficient methods have been developed to address this problem and are now routinely employed in molecular simulations~\cite{arnold_comparison_2013}. Nevertheless, when simulations scale to millions of degrees of freedom, as required for realistic modelling of complex biophysical and materials systems, long-range electrostatics continues to be one of the dominant computational bottlenecks~\cite{George:2022,ibeid:2020}. On modern heterogeneous many-core and GPU architectures, this cost is increasingly dominated by communication rather than arithmetic operations, motivating the development of algorithms with improved communication locality~\cite{Ayala:2022,Simmonett-2021,Zhou:2022}.

The separation of length scales in the electrostatic response is at the heart of state-of-the-art algorithms that treat the short- and long-range components with different numerical approaches. This strategy forms the basis of Ewald-type methods and their particle-mesh variants \cite{pollock:1996}. 
In Ewald summation, the Coulomb interaction is decomposed into a short-range contribution, evaluated by direct pair summation in real space, and a smooth long-range contribution, evaluated analytically in reciprocal space~\cite{ewald:1921}. 
Particle–mesh methods, including Particle Mesh Ewald~\cite{Darden:1993,Essmann:1995} and \ac{p3m}~\cite{Hockney:1981}, retain the Ewald short/long decomposition but accelerate the smooth long-range calculation by solving it numerically. In reciprocal space, the discretized Poisson equation becomes diagonal and can therefore be solved efficiently using three-dimensional \ac{fft}. The accuracy of these methods can be systematically controlled through the real-space cutoff, the Ewald splitting parameter, the mesh spacing, and the charge-assignment order, which together determine the balance between real-space truncation error and mesh/discretization error.
Particle–mesh methods combine systematic control of electrostatic accuracy with excellent computational efficiency and have consequently become the standard approach for large-scale molecular simulations. Although substantial effort has been devoted to improving the scalability of distributed \ac{fft} implementations in modern molecular simulation packages~\cite{Abraham:2015,Pall:2020,Phillips:2020}, the associated three-dimensional \ac{fft}s require global communication patterns that still limit scalability~\cite{Ayala:2022,ibeid:2020,Simmonett-2021}.
\\
Attempts to improve long-range solvers can be broadly divided into two classes: methods that avoid \ac{fft}s by adopting a different electrostatic formulation, and methods that retain the particle–mesh decomposition but replace the FFT-based mesh solver by a real-space one. The first class includes, for example, fast multipole and tree-based methods, that replace the particle-mesh solve by hierarchical expansions of distant charge distributions~\cite{Dehnen:2002,Gumerov:2004,Boateng-2019}. This can provide favorable scaling, but the accuracy is controlled by expansion order, clustering criteria, and the treatment of periodic images, rather than by the standard \ac{p3m} parameters of splitting, mesh spacing, and charge assignment. A different approach is provided by multilevel summation methods. These retain a separation between local interactions and smooth corrections, but the long-range contribution is distributed over a hierarchy of grids, leading to a different interpolation and error structure from a single \ac{p3m} mesh solve~\cite{Skeel:2002,Hardy:2014}. An alternative strategy is adopted by local-field methods, such as those introduced by Maggs and Rossetto, which reformulate electrostatics in terms of locally evolving electric fields satisfying Gauss's law~\cite{Maggs:2002,Rottler:2004}. These methods eliminate global \ac{fft}s but depart from the standard particle–mesh decomposition and therefore adopt a fundamentally different numerical framework. Finally, more local alternatives (e.g. damped or shifted pairwise approximations) modify the Coulomb potential itself, thus avoiding a global solve. The quality of these approximation depends, however, on the system and on the observables considered~\cite{Fennell:2006}.

For the present work, the relevant route is the second one: retaining the \ac{p3m} short/long decomposition and its associated error controls, while replacing the \ac{fft}-based mesh solve by a real-space solver. Within the class of \ac{p3m}-based methods, Beckers~\textit{et al.}\ employed successive over-relaxation to solve the discretized Poisson equation on a real-space grid~\cite{Beckers:1998}, while Sagui and Darden adopted a multigrid solver~\cite{Sagui:2001}, achieving linear scaling with the number of particles and grid points. Beckers \textit{et al.} noted that the converged solution from a previous configuration can serve as an effective initial guess for the subsequent iterative solve, but this observation was not developed further, and the multigrid approach of Sagui and Darden does not exploit an analogous initialization strategy. Related ideas have also been developed for non-periodic systems. In particular, Sutmann and Steffen proposed a particle--particle particle--multigrid method in which the near-field contribution is evaluated explicitly and the far-field contribution is obtained from a multigrid Poisson solver with boundary conditions determined by a multipole expansion~\cite{sutmann:2005}. These approaches demonstrate that replacing the \ac{fft}-based mesh solve is compatible with retaining the particle–mesh decomposition and its systematic accuracy control, motivating the search for more efficient real-space formulations. 

In this paper, we introduce \ac{p3maze}, a \ac{maze} formulation of particle--mesh electrostatics. The \ac{maze} formalism promotes appropriate auxiliary quantities, typically featuring as parameters subjected to conditions in the evolution of physical degrees of freedom, to zero-inertia dynamical variables whose values are determined by enforcing appropriate holonomic constraints during the dynamics. In this way, repeated solution of auxiliary optimization or field equations is replaced by constrained dynamical evolution while preserving the desired equilibrium distribution.

This approach was first proposed by Ciccotti and Ryckaert to describe rotation--translation coupling in diatomic molecules~\cite{Ryckaert:1981}. Over the past few years, it has been revived and generalized to a broad range of problems, including classical polarizable models~\cite{coretti:2018b,bonella:2020}, constant-potential electrochemical simulations~\cite{coretti:2020b}, shell models in magnetic fields~\cite{girardier:2021}, orbital-free first-principles \ac{md}~\cite{coretti:2022}, and advanced classical simulations of supercapacitors~\cite{angiolari:2025}. 

In very recent work~\cite{troni_mass-zero_2025}, the \ac{maze} formalism was applied to the solution of electrostatic interactions under the name \ac{pmaze}. In \ac{pmaze}, the electrostatic potential on a real-space grid is represented by zero-inertia auxiliary variables, while the discretized Poisson equation is enforced as a holonomic constraint. An efficient multigrid constraint solver yields linear scaling with the number of degrees of freedom and improves the efficiency of the iterative electrostatic solve. The formulation also preserves key physical properties, including energy and momentum conservation, time reversibility, and stationarity of ensemble averages. However, \ac{pmaze} treats the full Coulomb problem on the grid. Consequently, the fine spatial resolution required to represent the short-range electrostatic potential must be maintained throughout the simulation domain, increasing both the number of auxiliary variables and the size of the constrained problem.

The present work overcomes these limitations by combining the systematic accuracy framework of particle--mesh electrostatics with the \ac{maze} constrained-dynamics formalism. In \ac{p3maze}, the short-range contribution is evaluated by conventional particle--particle summation, while only the smooth long-range mesh potential is represented by zero-inertia auxiliary variables. The corresponding mesh equation is enforced as a holonomic constraint at every time step, replacing the conventional \ac{fft}-based reciprocal-space solve by a local constrained solver. The method preserves systematic control of accuracy. Beyond the formulation itself, we present a previously unrecognized property of the \ac{maze} dynamics: for the linear, time-independent Poisson operator considered here, the constrained correction problem is exactly equivalent to a direct Poisson solve initialized by a specific chronological extrapolation of the mesh potential. This result explains, for the first time, the origin of the accelerated convergence observed in the present and previous \ac{maze}-based electrostatic formulation, recasting what had previously been an empirical observation as a controlled feature of the formalism. Because the underlying extrapolation is independent of the constrained-dynamics formulation itself, this mechanism also provides a recipe for constructing (or rederiving) effective initial guesses for conventional iterative real-space Poisson solvers. We assess \ac{p3maze} by comparing the computed electrostatic potential, structural, and dynamical observables for molten NaCl and \ac{spc} water against reference \ac{pmaze}, \ac{p3m} and Ewald calculations, and by conducting a detailed analysis of computational performance and solver convergence.

\section{Theory}
\label{sec:theory}
Consider a system of $N_p$ point particles with coordinates $\{{\bm r_\beta}\}$, masses $\{m_\beta\}$, and charges $\{Q_\beta\}$. The corresponding microscopic charge density is

\begin{equation}
\rho(\bm r) = \sum_{\beta=1}^{N_p}Q_\beta \delta(\bm r-\bm r_\beta).
\end{equation}
Under periodic boundary conditions, the electrostatic energy includes the interaction of each charge with all of its periodic replicas and is given by

\begin{equation}
U_c
=
\frac{1}{2}
\sum_{\bm n\in\mathbb{Z}^3}
\sum_{\alpha=1}^{N_p}
\sum_{\beta=1}^{N_p}{}'
\frac{Q_\alpha Q_\beta}
{|\bm r_{\alpha\beta}+\bm n_{\rm cell}|},
\label{eq:coulomb_energy_periodic}
\end{equation}
where $\bm r_{\alpha\beta}=\bm r_\alpha-\bm r_\beta$, $\bm n_{\rm cell}=n_1\bm a_1+n_2\bm a_2+n_3\bm a_3$ is a lattice vector of the simulation cell, and the prime excludes the singular self interaction $\alpha=\beta$ for $\bm n_{\rm cell}=\bm0$. The lattice sum is conditionally convergent and is defined according to the usual Ewald convention~\cite{frenkelsmit2023}. Following the standard particle--mesh formulation~\cite{Hockney:1981}, the periodic Coulomb interaction is decomposed into a rapidly decaying short-range contribution and a smooth long-range contribution by introducing a screening charge density associated with each particle. The total screening density is

\begin{equation}
\rho^s(\bm r)
=
\sum_{\beta=1}^{N_p}
\rho_\beta^s(\bm r;\{{\bm r_\beta}\}),
\label{eq:smooth_density}
\end{equation}
In the present work we adopt Gaussian screening functions,
\begin{equation}
\rho^s_\beta(\bm r;\{{\bm r_\beta}\})
=
\frac{Q_\beta}
{(2\pi\sigma^2)^{3/2}}
\exp\left[
-\frac{|\bm r-\bm r_\beta|^2}
{2\sigma^2}
\right],
\label{eq:gaussian_screening_density}
\end{equation}
consistent with the formulation of Beckers~\cite{Beckers:1998}.
The electrostatic energy is written, exactly, as
\begin{equation}
U_c
=
U_{\rm SR}
+
U_{\rm LR}
+
U_{\rm self}.
\label{eq:p3m_decomposition}
\end{equation}
With our choice for the screening density, the short-range contribution - given by the difference between the Coulomb interaction of the point charges and that of the corresponding screening densities - is,
\begin{equation}
\begin{split}
U_{\rm SR}
&
=
\frac{1}{2}
\sum_{\bm n\in\mathbb Z^3}
\sum_{\alpha=1}^{N_p}
\sum_{\beta=1}^{N_p}{}'
\left[
\frac{Q_\alpha Q_\beta}
{|\bm r_{\alpha\beta}+\bm n_{\rm cell}|}
-
\int d\bm r d\bm r'
\frac{\rho_\alpha^s(\bm r;\{{\bm r_\alpha}\})\rho_\beta^s(\bm r';\{{\bm r_\beta}\})}
{|\bm r-\bm r'+\bm n_{\rm cell}|}
\right]
\simeq
\\
&
\simeq
\frac{1}{2}
\sum_{\alpha=1}^{N_p}
\sum_{\beta\neq\alpha}^{r_{\alpha\beta}<R_c}
\frac{Q_\alpha Q_\beta}
{r_{\alpha\beta}}
\left[
1-
\operatorname{erf}
\left(
\frac{r_{\alpha\beta}}
{\sqrt{2}\sigma}
\right)
\right].
\label{eq:p3maze_vsr_general}
\end{split}
\end{equation}
In the second line, the rapidly decaying short-range contribution is evaluated using the minimum-image convention and truncated at the real-space cutoff $R_c$. For the cubic cells considered here, $R_c\leq L/2$. The magnitude of the contribution in the expression above, i.e. the short-range tail beyond $R_c$, is controlled by the choice of $R_c$ and $\sigma$. The term
\begin{equation}
\begin{split}
  U_{\rm self}
&=
-\frac{1}{2}
\sum_{\alpha=1}^{N_p}
\int d\bm r d\bm r'
\frac{\rho_\alpha^s(\bm r;\{{\bm r_\alpha}\})\rho_\alpha^s(\bm r';\{{\bm r_\alpha}\})}
{|\bm r-\bm r'|}=
\\
&=
-\frac{1}{\sqrt{2\pi}\sigma}
\sum_{\alpha=1}^{N_p}
Q_\alpha^2.
\label{eq:p3maze_vself_general}  
\end{split}
\end{equation}
removes the self interaction of each screening density.

The smooth long-range interaction is

\begin{equation}
U_{\rm LR}
=
\frac{1}{2}
\sum_{\bm n\in \mathbb{Z}^3}
\sum_{\alpha=1}^{N_p}
\sum_{\beta=1}^{N_p}
\int d\bm r d\bm r'
\frac{\rho_\alpha^s(\bm r;\{{\bm r_\alpha}\})\rho_\beta^s(\bm r';\{{\bm r_\beta}\})}
{|\bm r-\bm r'+\bm n_{\rm cell}|},
\label{eq:p3maze_vlr_general}
\end{equation}

which retains the full periodic long-range electrostatic contribution. Rather than evaluating Eq.~\eqref{eq:p3maze_vlr_general} directly, particle--mesh methods determine the corresponding electrostatic potential by solving the Poisson equation
\begin{equation}
\nabla^2\phi_{\rm LR}(\bm r)
=
-4\pi
\rho^s(\bm r),
\label{eq:poisson_eq}
\end{equation}
which, after discretization on a uniform grid of spacing $h$, becomes

\begin{equation}
\bm M\bm\phi_{\rm LR}=-4\pi\frac{\bm q^s}{h},
\label{eq:p3maze_discrete_poisson}
\end{equation}
where $\bm M/h^2$ is the discrete Laplacian and $\bm q^s/h^3$ is the vector of assigned smooth charges. Under periodic boundary conditions, the discrete Laplacian $\bm M$ has a constant null mode, so the electrostatic potential is defined only up to an additive constant. We fix this freedom by imposing the gauge $\langle\bm{\phi}_{\rm LR}\rangle=0$ after each field update. Within the constrained-dynamics formulation described below, this gauge condition can equivalently be interpreted as an additional linear constraint selecting a unique representative of the mesh potential.

Up to this point, the formulation of the problem follows the standard particle--mesh formulation. The remaining numerical task is the solution of the discrete Poisson equation at every simulation step. Conventional particle--mesh methods perform this solve using either FFT-based reciprocal-space techniques or iterative real-space solvers.

The novel idea that \ac{p3maze} brings in is to reformulate the long-range particle--mesh problem as a constrained dynamical system. Instead of solving the discrete Poisson equation Eq.~\eqref{eq:p3maze_discrete_poisson} at every step along a \ac{md} trajectory, the values of the long-range electrostatic potential on the mesh are promoted to auxiliary dynamical variables. The discrete Poisson equation is then enforced as a holonomic constraint within the \ac{maze} formalism, following the approach previously developed for \ac{pmaze}~\cite{troni_mass-zero_2025}.

To this end, the components of the discretized long-range electrostatic potential, ${\bm\phi_{\rm LR}}$, are assigned a finite auxiliary inertia $\mu$ (whose vanishing limit will be considered below), leading to the extended Lagrangian

\begin{equation}
\begin{split}
\mathcal L_{\rm P3MaZe}
=
&
\frac{1}{2}
\sum_{\beta=1}^{N_p}
m_\beta
\dot{\bm r}_\beta^T
\dot{\bm r}_\beta
+
\frac{1}{2}
\mu
\dot{\bm\phi}_{\rm LR}^T
\dot{\bm\phi}_{\rm LR}
\\
&
-
\bar U(\{\bm r_\beta\})
-
U_{\rm SR}(\{\bm r_\beta\})
-
U_{\rm LR}(\bm\phi_{\rm LR};\{\bm r_\beta\})
-
U_{\rm self}
-
\bm\lambda^T
\bm\sigma_{\rm LR}
\left(
\bm\phi_{\rm LR};\{\bm r_\beta\}
\right),
\end{split}
\label{eq:p3maze_lagrangian}
\end{equation}
where $\bar U(\{\bm r_\beta\})$ contains all non-electrostatic interactions and $\bm\lambda$ denotes the vector of Lagrange multipliers associated with the constraints. The discretized long-range electrostatic energy is

\begin{equation}
U_{\rm LR}\left(\bm\phi_{\rm LR};\{\bm r_\beta\}\right)
=
\frac{h}{8\pi}
\bm\phi_{\rm LR}^T
\bm M
\bm\phi_{\rm LR}
+
\bm\phi_{\rm LR}^T
\bm q^s,
\label{eq:p3maze_vlr_discrete}
\end{equation}
and the holonomic constraints are defined as the stationarity conditions of the long-range energy,

\begin{equation}
\begin{split}
\bm\sigma_{\rm LR}
\left(
\bm\phi_{\rm LR};\{\bm r_\beta\}
\right)
&=
\nabla_{\bm\phi_{\rm LR}^T}
U_{\rm LR}
\left(
\bm\phi_{\rm LR};\{\bm r_\beta\}
\right)=
\\
&=
\frac{h}{4\pi}
\bm M
\bm\phi_{\rm LR}
+
\bm q^s(\{\bm r_\beta\})
=
\bm0.
\label{eq:p3maze_constraint}
\end{split}
\end{equation}
Eq.~\eqref{eq:p3maze_constraint} is the discrete Poisson equation (Eq.~\eqref{eq:p3maze_discrete_poisson}) expressed as a holonomic constraint. Consequently, the constraint manifold coincides with the set of mesh potentials satisfying the discrete Poisson equation. Furthermore, when the constraint is satisfied,

\begin{equation}
U_{\rm LR}
=
\frac12
\bm\phi_{\rm LR}^T
\bm q^s,
\label{eq:p3maze_vlr_on_constraint}
\end{equation}
which is the standard discrete expression for the electrostatic energy of the smooth charge distribution~\cite{note:discrete_energy}.
% \bibnote{Eq.~\eqref{eq:p3maze_vlr_on_constraint} is the discrete counterpart of $U_{\rm LR}=\frac12\int_\Omega d\bm r,\rho^s(\bm r)\phi_{\rm LR}(\bm r)$, equivalent to the Green-function representation in Eq.~\eqref{eq:p3maze_vlr_general}.}

The evolution equations for the extended system follow from the Euler--Lagrange equations. As in the general \ac{maze} formalism~\cite{coretti:2018b,troni_mass-zero_2025}, the physical dynamics are recovered in the limit $\mu\rightarrow 0$. In this limit, the Lagrange multipliers scale linearly with $\mu$, ensuring that the acceleration of the auxiliary variables remains finite. As shown in Section 1 of the Supporting Information, the resulting equations of motion are

\begin{equation}
\label{eq:MaZeEvolution}
\begin{split}
m_\alpha \ddot{\bm r}_\alpha
&=
-\nabla_{\bm{r}_\alpha}{U_{\rm SR}}(\{\bm{r}_\beta\})-  \nabla_{{\bm r}_\alpha}U_{\rm LR}(\bm{\phi};\{\bm{r}_\beta\})  -\nabla_{\bm{r}_\alpha}{\bar U}(\{\bm{r}_\beta\}),
\\
\ddot{\bm\phi}_{\rm LR}
&=
-\bm M\bm\eta,
\end{split}
\end{equation}

where $\bm\eta=\lim_{\mu\rightarrow0}
\frac{h}{4\pi}
\frac{\bm\lambda}{\mu}
$ is the rescaled vector of Lagrange multipliers.

In practice, the physical degrees of freedom are propagated using the velocity--Verlet algorithm, while the auxiliary variables are integrated using the Verlet scheme. At each time step, the rescaled Lagrange multipliers are determined by enforcing the constraint $\bm\sigma_{\rm LR}
    \left(
    \bm\phi_{\rm LR};\{\bm r_\beta\}
    \right)=\bm 0$
, via an adapted SHAKE algorithm~\cite{ryckaert:1977}. Thus, the dynamics evolve directly on the constraint manifold defined by Eq.~\eqref{eq:p3maze_constraint}, ensuring that the mesh potential remains consistent with the smooth charge density throughout the simulation. 

\paragraph{\ac{p3maze} and \ac{p3m}: Constrained propagation and temporal extrapolation.}
\vspace{3mm}

We now show that, for the linear (in the auxiliary variables) Poisson constraint considered here, the solution of the \ac{p3maze} constraint problem and the direct real-space Poisson solve are algebraically equivalent, and that this equivalence gives the constrained dynamics a previously unrecognized interpretation as a chronological predictor for the mesh potential obtained via standard \ac{p3m} solvers. As mentioned above, in a conventional real-space \ac{p3m} approach, the long-range potential is obtained by directly solving the discretized Poisson equation at each \ac{md} step using a chosen initial guess. In \ac{p3maze}, the same solution is reached instead by propagating the auxiliary mesh potential and correcting it so as to satisfy the Poisson constraint. At time step $k+1$, the full update of the mesh potential in \ac{p3maze} is given by a standard Verlet step
\begin{equation}
\bm\phi_{\rm LR}^{k+1}=\bm\phi_{{\rm LR},p}^{k+1}-\frac{4\pi}{h}\bm y^{k+1},
\label{eq:phi_update}
\end{equation}
where $\bm\phi_{{\rm LR},p}^{k+1}$ denotes the provisional potential predicted by the Verlet propagation, $\bm\phi_{{\rm LR},p}^{k+1}=2\bm\phi_{\rm LR}^{k}-\bm\phi_{\rm LR}^{k-1}$, and $\bm y^{k+1}$ is the SHAKE correction that enforces the constraint. As detailed in the Supporting Information, the latter is related to the rescaled Lagrange multipliers through
\begin{equation}
\bm y^{k+1}=\frac{h}{4\pi}\Delta t^2\bm M\bm\eta^{k+1}.
\label{eq:correction_definition}
\end{equation}

The constraint dynamics then focuses on obtaining the unknown quantity ${\bm y}^{k+1}$ rather than directly the potential. These corrections are determined via the SHAKE condition
\begin{equation}
\label{eq:discPoisson}
\bm\sigma_{\rm LR}\left(\bm\phi_{\rm LR}^{k+1};\{\bm r_\beta^{k+1}\}\right)=\bm 0 \ ,
\end{equation}

by substituting the Verlet prediction for the potential in the argument
\begin{equation}
\bm\sigma_{\rm LR}\left(\bm\phi_{{\rm LR},p}^{k+1}-\frac{4\pi}{h}\bm y^{k+1};\{\bm r_\beta^{k+1}\}\right)=\bm 0.
\end{equation}
The equation above can be trivially rewritten as
\begin{equation}
\bm M\bm y^{k+1}=\frac{h}{4\pi}\bm M \bm\phi_{{\rm LR},p}^{k+1} + \bm{q}^{s,k+1} \equiv \bm\sigma_{{\rm LR},p}^{k+1}.
\label{eq:correction_system}
\end{equation}
where $\bm\sigma_{{\rm LR},p}^{k+1}$ is the value of the constraint computed at provisional potential. 
This argument shows that the discrete Poisson equation for the potential at step $k+1$, Eq.~\eqref{eq:discPoisson}, and the SHAKE solution for $\bm y^{k+1}$ are algebraically equivalent formulations of the linear system for the long-range electrostatic potential. The first is typically solved iteratively after input of the initial guess for the potential. The second is also solved iteratively, possibly - as done later in this work - with the same solver used for standard Poisson - but can be conveniently interpreted as a predictor-corrector scheme in which the predictor is the provisional potential and the correction with the SHAKE step.  

An interesting consequence of the discussion above pertains to the initialization of the iterative solve. In \ac{p3maze}, the initial guess is specified for the correction variable $\bm y$ not directly for the potential. We denote by $\bm y_0^{(m)}$ a predictor constructed from $m$ previous converged corrections, with $\bm y_0^{(0)}=\bm 0$. Through Eq.~\eqref{eq:phi_update}, each choice of $\bm y_0^{(m)}$ determines a corresponding initial estimate of the mesh potential that can be fed to the direct solver.
We denote by $\bm\phi_{{\rm LR},0}^{(m)}$ the initial mesh-potential estimate corresponding to the correction predictor $\bm y_0^{(m)}$.

For $\bm y_0^{(0)}=\bm 0$, the initial potential is simply the provisional Verlet prediction,
\begin{equation}
\label{eq:potVerletinit}
\bm\phi_{{\rm LR},0}^{(0)}=\bm\phi_{{\rm LR},p}^{k+1}=2\bm\phi_{\rm LR}^{k}-\bm\phi_{\rm LR}^{k-1}.
\end{equation}
A more informative initialization is obtained by retaining the converged correction from the preceding time step. 
For the standard \ac{p3maze} choice $\bm y_0^{(1)}=\bm y^k$, the update relation at the preceding time step gives
\begin{equation}
\bm\phi_{{\rm LR},0}^{(1)}=\bm\phi_{{\rm LR},p}^{k+1}-\frac{4\pi}{h}\bm y^k=3\bm\phi_{\rm LR}^{k}-3\bm\phi_{\rm LR}^{k-1}+\bm\phi_{\rm LR}^{k-2}.
\end{equation}

Thus, reusing the previous converged correction is exactly equivalent, for the fixed linear operator considered here, to a quadratic chronological extrapolation of the converged mesh potential.

More generally, for the uniform time step used here, higher-order predictors can be constructed from the $m$ most recent converged corrections as
\begin{equation}
\bm y_0^{(m)}=\sum_{j=0}^{m-1}(-1)^j\binom{m}{j+1}\bm y^{k-j},\qquad m\geq1.
\label{eq:correction_predictor}
\end{equation}

Using Eq.~\eqref{eq:phi_update} and the corresponding relations at the preceding time steps, this gives
\begin{equation}
\bm\phi_{{\rm LR},0}^{(m)}=\sum_{j=0}^{m+1}(-1)^j\binom{m+2}{j+1}\bm\phi_{\rm LR}^{k-j}.
\label{eq:chronological_phi_predictor}
\end{equation}
Thus, the constrained propagation provides an explicit correspondence between predictors for the correction variable and chronological predictors for the mesh potential.  Eq.~\eqref{eq:chronological_phi_predictor} is not strictly dependent on its dynamical origin and can therefore be employed to initialize different iterative solvers for real space electrostatics.

In general, temporal extrapolation of previous solutions, including Lagrange polynomial predictors, is a well-established strategy for accelerating sequences of iterative linear solves~\cite{austin_initial_2020}. In \ac{p3maze}, the extrapolation mechanism arises naturally from the constrained propagation of the auxiliary mesh potential, providing a direct dynamical basis for the higher-order predictors introduced above.
In previous work, Beckers \textit{et al.}~\cite{Beckers:1998} noted that the converged electrostatic potential from the previous ionic configuration can provide an effective initial guess for the subsequent solve. However, attempts to improve on this via a Taylor expansion in time of the potential did not prove fruitful for the systems they considered and was not pursued further. Connected strategies have also been developed in molecular simulation methods involving auxiliary or self-consistent variables but not necessarily transferred to electrostatics in classical MD. For example, K\"ostler \textit{et al.} used previous Hartree potentials together with - only - second order temporal extrapolation to initialize successive Poisson solves in first principles Car Parrinello \ac{md}~\cite{kostler_parallel_2008}. For the simulation of classical polarizable models, Kolafa introduced a predictor--corrector scheme based on the history of self-consistent polarization variables~\cite{kolafa:2004}. The Kolafa scheme has been adapted also to first principle MD in the so-called second-generation Car Parrinello dynamics~\cite{kuhne:2014}. However, the predictor--corrector construction of Kolafa requires to tune system dependent parameters and is not fully time-reversible. In contrast, because the \ac{p3maze} predictor-corrector scheme is embedded in the constrained dynamics of the system, it automatically inherits convenient properties such as symplecticity and time-reversibility.

\paragraph{\ac{p3maze} and \ac{pmaze}: Charge assignment and smoothing.}

An important distinction between \ac{pmaze} and \ac{p3maze} concerns the source term of the mesh Poisson equation. In \ac{pmaze}, the full particle charge density is assigned directly to the grid and acts as the source of the discretized Poisson problem. In \ac{p3maze}, by contrast, the source term is the smooth screening charge density arising from the particle--mesh decomposition, as in conventional real-space \ac{p3m} methods~\cite{Beckers:1998,Sagui:2001}. Consequently, \ac{p3maze} inherits the systematic accuracy framework of particle--mesh electrostatics.
In the present implementation, particle charges are first assigned to the mesh using cubic B-splines. The resulting charge density is subsequently smoothed over a finite stencil to obtain the mesh density $\bm q^s$ entering Eq.~\eqref{eq:p3maze_constraint}. This two-step construction provides the discrete particle--mesh representation of the continuous Gaussian screening density introduced above. The B-spline assignment is part of the mesh discretization, rather than an additional physical smoothing, and its effect is consistently accounted for by using the same B-spline weights when interpolating the mesh field back to the particles before computing the force. The resulting particle-level force therefore represents the long-range force generated by the screened Gaussian charge density within the adopted mesh discretization, as discussed in the paragraph on electrostatic forces. Because the source term is smooth, the corresponding mesh potential is also smoother than in \ac{pmaze}, with reduced high-frequency components of the mesh potential. This in turn reduces the mesh resolution required to achieve a given target accuracy and therefore the number of necessary auxiliary variables. This additional smoothing operation introduces an extra particle--mesh cost that is absent in \ac{pmaze}. As shown in the Results, however, this overhead is more than compensated by the reduced cost of the solver, leading to an overall improvement in computational efficiency.

\paragraph{\ac{p3maze} and \ac{pmaze}: Electrostatic energy.}
In \ac{pmaze}, the electrostatic potential is obtained from the full discretized Poisson problem. Consequently, a direct evaluation of the mesh energy contains a grid self-energy associated with the discrete representation of each particle charge~\cite{Hockney:1981,troni_mass-zero_2025}. This contribution depends on both the charge-assignment scheme and the mesh resolution, making its removal numerically delicate. For this reason, \ac{pmaze} evaluates changes in the electrostatic energy from the mechanical work $U_c^{\rm work}$ performed by the electrostatic forces along the trajectory,
\begin{equation}
U_c^{\rm work}(t)-U_c^{\rm work}(0)
=
-\sum_{\alpha=1}^{N_p}
\int_{\bm r_\alpha^0}^{\bm r_\alpha(t)}
\bm F_\alpha\cdot d\bm r_\alpha,
\label{eq:electrostatic_work}
\end{equation}
where $U_c^{\rm work}(0)$ fixes the arbitrary energy reference. In \ac{p3maze}, the singular Coulomb interaction is treated analytically through $U_{\rm SR}$, while the mesh represents only the smoothed long-range contribution $U_{\rm LR}$. Consequently, the grid self-energy associated with discretized point charges is absent, and the only remaining self contribution is the analytical Gaussian screening correction given by Eq.~\eqref{eq:p3maze_vself_general}. The smoothing procedure provides a well-behaved explicit estimate of the long-range energy $U_{\rm LR}$ (Eq.~\eqref{eq:p3maze_vlr_on_constraint}). 

At finite mesh resolution, the work-derived energy is more consistent with the computed forces than the explicit mesh energy and is therefore used in the NVE analysis. However, by the work–energy theorem, changes in kinetic energy are compensated, up to time-integration and work-quadrature errors, by opposite changes in the energy reconstructed from those forces. The conservation of the resulting total energy is therefore primarily a consistency test between force propagation and work integration, rather than an independent validation of the dynamics. Independent stability tests are instead provided by kinetic energy or temperature drift and by their dependence on the time step and solver tolerance. The comparison between the two energy definitions and the NVE energy tests are reported in Section~2 of the Supporting Information.

\paragraph{Electrostatic forces.}

The electrostatic force follows the same decomposition as the energy,

\begin{equation}
\bm F_\alpha
=
\bm F_\alpha^{\rm SR}
+
\bm F_\alpha^{\rm LR},
\label{eq:force_decomposition}
\end{equation}
where the short-range contribution is evaluated analytically. For the Gaussian screening function (Eq.~\eqref{eq:gaussian_screening_density}),

\begin{equation}
\bm F^{\rm SR}_\alpha
=
\sum_{\beta\ne\alpha}^{r_{\alpha\beta}<R_c}
Q_\alpha Q_\beta
\left[
\frac{\operatorname{erfc}\left(r_{\alpha\beta}/\sqrt{2}\sigma\right)}
{r_{\alpha\beta}^{3}}
+
\sqrt{\frac{2}{\pi}}
\frac{\exp\left[-r_{\alpha\beta}^{2}/(2\sigma^2)\right]}
{\sigma r_{\alpha\beta}^{2}}
\right]
\bm r_{\alpha\beta},
\label{eq:short_range_force}
\end{equation}
where $\bm r_{\alpha\beta}=\bm r_\alpha-\bm r_\beta, (r_{\alpha\beta}=|\bm r_{\alpha\beta}|$), and the sum is restricted to particle pairs within the real-space cutoff $R_c$.

The long-range contribution is obtained from the constrained mesh potential using the same interpolation procedure employed in \ac{pmaze}. For example, the (x)-component is

\begin{equation}
F_{\alpha,x}^{\rm LR}
=
-Q_\alpha
\sum_{m}
W(\bm r_\alpha-\bm r_m)
D_x\phi_{{\rm LR},m},
\label{eq:long_range_force}
\end{equation}
where $D_x$ denotes the finite-difference approximation to the mesh gradient and $W$ is the particle--mesh interpolation function. Greek subscripts label particles, whereas Latin subscripts denote mesh points. In the present implementation, $W$ is a cubic B-spline, identical to the charge-assignment function. Using the same interpolation function for charge assignment and force reconstruction ensures consistency between the two operations, as in \ac{pmaze}.

The \ac{p3maze} force construction preserves the momentum-conserving properties of both components of the particle--mesh decomposition. The short-range contribution is pairwise and antisymmetric and therefore conserves total momentum exactly (up to floating-point roundoff). As in \ac{pmaze}, the long-range contribution conserves momentum on the mesh provided that charge assignment and force interpolation employ the same weight function and the Laplacian is discretized using central finite differences~\cite{Hockney:1981}. In the present implementation, both conditions are satisfied by using cubic B-splines together with the standard seven-point finite-difference stencil. Higher-order Hermitian discretizations~\cite{Sagui:2001} could be adopted straightforwardly if greater mesh accuracy were required.

Taken together, these differences show that \ac{p3maze} combines the systematic accuracy and analytical structure of particle--mesh electrostatics with the constrained-dynamics formulation and efficient iterative correction strategy of \ac{pmaze}. Momentum conservation tests and further discussion on the topic can be found in Section 2 of the Supporting information.

\section{Results}
\label{sec:Results}
\subsection{Molten salt}
\label{subsec:moltensalt}

\subsubsection{System and simulation protocol}
As a first validation of the method, we consider molten NaCl, which provides a direct comparison with our previous \ac{pmaze} formulation~\cite{troni_mass-zero_2025}. Since \ac{p3maze} differs from \ac{pmaze} only in the treatment of the long-range electrostatic problem, this system allows us to isolate the effect of introducing the particle--mesh decomposition while keeping the underlying constrained-dynamics framework unchanged. In a later section, and in the Supplementary Information, we also report results on the accuracy of the electrostatic potential computed with \ac{p3maze} compared to a standard P3M solver. 

The simulated system consists of 250 ions (125 Na$^+$ and 125 Cl$^-$) in a cubic simulation cell of side $L=20.64~\text{\AA}$, corresponding to a density of $1.3793~\mathrm{g \cdot cm^{-3}}$. Periodic boundary conditions are applied in all directions. Non-electrostatic interactions are described by the Born--Huggins--Meyer potential with Tosi--Fumi parameters, following the model adopted in previous simulations of molten NaCl~\cite{Galamba:2007,Mouhat:2013,troni_mass-zero_2025}.

Initial configurations were generated from a bcc lattice with Maxwell--Boltzmann velocities at $T=1550$ K. After equilibration in the canonical ensemble using Langevin dynamics integrated with the OVRVO splitting scheme~\cite{Sivak:2013}, production trajectories were generated in the microcanonical ensemble by setting the friction coefficient to zero, thereby recovering the velocity--Verlet limit. The equilibration stage lasted $2.5$ ps with a time step of $0.25$ fs. For \ac{p3maze}, the electrostatic parameters were chosen according to the convergence analysis reported in Section 3 of the Supporting Information. The Gaussian screening width was set to $\sigma=1.39~~\text{\AA}$, corresponding to an Ewald splitting parameter $\beta=0.5087~\text{\AA}^{-1}$, the real-space cutoff was $R_c=4.5~\text{\AA}$, and the mesh consisted of $60^3$ grid points ($h=0.344~\text{\AA}$). The smooth charge density was generated using the diffusion-based smoothing procedure of Beckers and co-workers~\cite{Beckers:1998,Sagui:2001}, whose computational cost scales linearly with the number of mesh points. The number of diffusion steps $N_{\mathrm{diff}}$ is selected so that the variance of the discrete diffusion kernel reproduces the target Gaussian width $\sigma$: $N_{\mathrm{diff}}=\operatorname{round}[(\sigma/h)^2/(2D_g)]$ where $h$ is the mesh spacing and $D_g=1/6.2$ is the dimensionless diffusion coefficient~\cite{Sagui:2001}. For comparison, \ac{pmaze} calculations employed the optimized setup of Ref.~\cite{troni_mass-zero_2025}, using a $120^3$ mesh ($h=0.172~\text{\AA}$). In both methods, the constraint equations were solved at every molecular dynamics step to a tolerance of $10^{-7}$.
\subsubsection{Structural and dynamical validation}

The purpose of the following analysis is to establish that replacing the full-grid Poisson formulation of \ac{pmaze} by the particle--mesh decomposition of \ac{p3maze} preserves both equilibrium and dynamical properties. We therefore compare structural, single-particle transport, and collective transport observables obtained with the two methods against available literature data.

\begin{figure}[h!]
    \centering
    \includegraphics[width=0.65\linewidth]{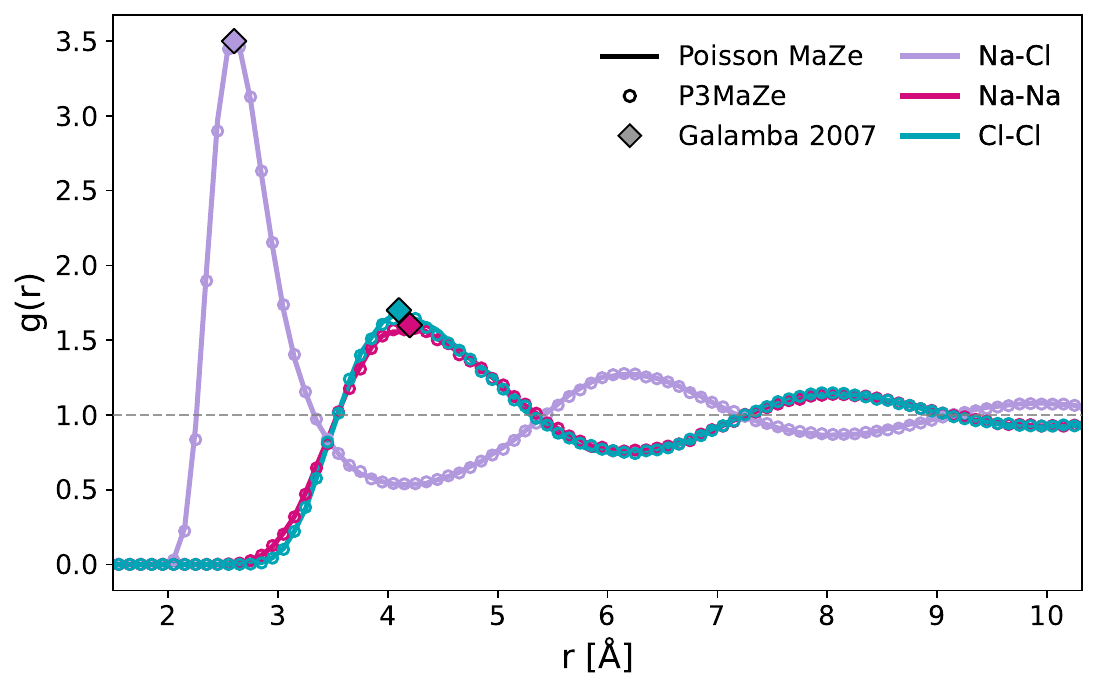}
    \caption{Partial radial distribution functions of molten NaCl obtained with \ac{pmaze} and \ac{p3maze}. The Na--Cl, Na--Na and Cl--Cl distributions are compared with the reference peak positions reported by Galamba \textit{et al.}~\cite{Galamba:2007}. The two methods are in quantitative agreement, indicating that the particle--mesh decomposition preserves the equilibrium liquid structure.}
    \label{fig:g_r_moltensalt}
\end{figure}

The equilibrium liquid structure is first assessed through the partial radial distribution functions. Fig. ~\ref{fig:g_r_moltensalt} compares the Na--Cl, Na--Na and Cl--Cl distributions obtained from 25 ps microcanonical trajectories with those reported by Galamba \textit{et al.}~\cite{Galamba:2007}. The \ac{p3maze} and \ac{pmaze} curves are indistinguishable within graphical resolution for all ionic pairs, demonstrating that introducing the particle--mesh decomposition does not modify the equilibrium structure of the liquid.

Dynamical observables provide a more stringent validation because they accumulate the effect of the electrostatic forces over extended trajectories. We first examine the \ac{vacf}, from which the self-diffusion coefficients are obtained as
\begin{equation}
    D_a = \frac{1}{3N_a}\sum_{\alpha \in a}\int_0^\infty dt\,
    \left\langle \bm v_\alpha^a(t)\cdot \bm v_\alpha^a(0) \right\rangle ,
    \label{eq:self_diffusion_vacf}
\end{equation}
where \(a\) denotes the ionic species and the average is taken over the $N_a$ particles of that species and over time origins.\\

\begin{figure}[h!]
    \centering
    \includegraphics[width=0.99\linewidth]{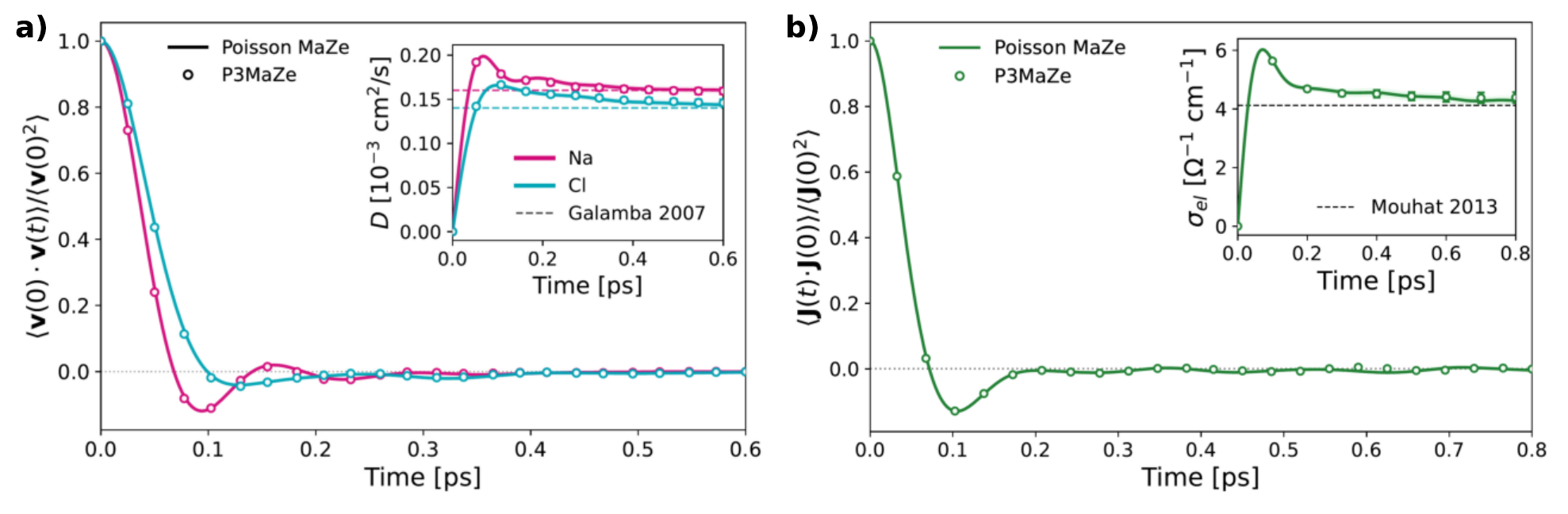}
    \caption{Comparison of single-particle and collective dynamical observables for molten NaCl obtained with \ac{pmaze} and \ac{p3maze}. \textbf{a)} \ac{vacf} of Na$^+$ and Cl$^-$ ions together with the corresponding cumulative Green--Kubo integrals used to determine the self-diffusion coefficients. \textbf{b)} \ac{cacf} and cumulative Green--Kubo integrals yielding the electrical conductivity. Shaded regions indicate the standard error estimated from independent trajectories. The two methods yield statistically indistinguishable translational dynamics and finite-time molecular charge responses, demonstrating that introducing the particle--mesh decomposition into the \ac{maze} framework preserves both single-particle and collective dynamical properties.}
    \label{fig:vacf_caf_moltensalt}
\end{figure}

Diffusion coefficients were estimated from ensembles of ten independent 15 ps microcanonical trajectories for each electrostatic method. Statistical uncertainties correspond to the standard error of the replica averages. Fig.~\ref{fig:vacf_caf_moltensalt}\textbf{a)} compares the normalized \ac{vacf}s together with the cumulative diffusion coefficients. Agreement between \ac{pmaze} and \ac{p3maze} is observed over the entire correlation time, leading to statistically indistinguishable diffusion coefficients:
$D_{\mathrm{Na}}=(0.161\pm0.004)\times10^{-3} \ \mathrm{cm^2\cdot s^{-1}}$ and
$D_{\mathrm{Cl}}=(0.144\pm0.004)\times10^{-3} \ \mathrm{cm^2 \cdot s^{-1}}$ for \ac{pmaze},
compared with
$D_{\mathrm{Na}}=(0.159\pm0.004)\times10^{-3} \ \mathrm{cm^2\cdot s^{-1}}$ and
$D_{\mathrm{Cl}}=(0.146\pm0.004)\times10^{-3} \ \mathrm{cm^2\cdot s^{-1}}$ for P3MaZe.
Both methods also agree with the reference values of Galamba \textit{et al.}~\cite{Galamba:2007}.

As a more demanding test, we next consider the \ac{cacf}, with the current computed as $\mathbf{J}(t) = \sum^{N_p}_{\alpha=1} Q_\alpha \mathbf{v}_\alpha(t)$. This correlation probes collective charge transport and is therefore particularly sensitive to the treatment of long-range electrostatic interactions~\cite{Beckers:1998}. The corresponding electrical conductivity was obtained from 
\begin{equation}
    \sigma_{\rm el}(t) =
    \frac{1}{3 V k_{\rm B} T}
    \int_0^t
    \left\langle
    \mathbf{J}(0)\cdot\mathbf{J}(t')
    \right\rangle
    dt',
\end{equation}
using ensembles of twenty independent 40 ps trajectories. In the equation above $V$ is the volume of the system. Fig.~\ref{fig:vacf_caf_moltensalt} \textbf{b)} shows that the current autocorrelation functions and cumulative conductivities obtained with \ac{pmaze} and \ac{p3maze} are statistically indistinguishable. The resulting conductivities,
$\sigma_{\rm el}=4.29\pm0.15~\Omega^{-1}\cdot \mathrm{cm}^{-1}$ for \ac{pmaze} and
$\sigma_{\rm el}=4.39\pm0.18~\Omega^{-1}\cdot \mathrm{cm}^{-1}$ for \ac{p3maze},
are mutually consistent and remain close to the reference value reported by Mouhat \textit{et al.}~\cite{Mouhat:2013}.

These structural and dynamical results demonstrate that introducing the particle--mesh decomposition into the \ac{maze} framework leaves both equilibrium and transport properties unchanged within statistical uncertainty. The \ac{p3m} decomposition therefore preserves the physical accuracy of the original constrained formulation while substantially reducing the computational cost of the long-range electrostatic treatment, as quantified in the performance analysis shown later in this work.

\subsection{\ac{spc} model}
Having established that \ac{p3maze} reproduces the original \ac{pmaze} formulation for a strongly ionic liquid, we consider a substantially more demanding benchmark based on the flexible \ac{spc} water model introduced by Wu, Tepper, and Voth~\cite{wu:2006}. Unlike molten salts, liquid water is not dominated by strong ionic correlations and high screening effects. Long-range electrostatic interactions instead play a central role in determining the hydrogen-bond network, molecular reorientation, and transport properties, making water a particularly sensitive test of the accuracy of the electrostatic treatment.

\ac{spc} is a flexible three-site model in which the interaction sites coincide with the atomic nuclei. Each water molecule is described by two O--H bond lengths and one H--O--H bond angle (all represented as harmonic interactions), allowing the intramolecular degrees of freedom to couple continuously to the electrostatic environment. The parameters of the model were taken from Ref.~\cite{wu:2006} and are summarized in the Supporting Information. As in the original \ac{spc} model, Lennard--Jones interactions act only on the oxygen site.

\subsubsection{System and simulation protocol}

In contrast with the molten-salt benchmark, where the comparison isolates the effect of introducing the particle--mesh decomposition into the \ac{maze} framework, the \ac{spc} system assesses whether \ac{p3maze} reproduces the structural and dynamical properties obtained with established particle--mesh electrostatic methods. The primary comparison is performed against the \ac{p3m} implementation in LAMMPS~\cite{LAMMPS} through time-resolved structural and dynamical observables. As an additional consistency check, integrated observables are also compared with Ewald summation and the original \ac{pmaze} formulation (Table~\ref{tab:water_spcfw_observables}).

The \ac{p3m} and Ewald calculations were performed with LAMMPS, whereas \ac{p3maze} and \ac{pmaze} employed our in-house implementation. All four methods were applied to identical molecular configurations and simulation conditions.
The simulated system consists of 216 water molecules in a cubic periodic box of side $L=18.552~\text{\AA}$, corresponding to a density of $1.012~\mathrm{g \cdot cm^{-3}}$, at $T=298.15~\mathrm{K}$. After a $1$ ns equilibration in the canonical ensemble using a time step of $0.5~\mathrm{fs}$, production data were collected from forty distinct $150~\mathrm{ps}$ microcanonical trajectories. Initial configurations for the production runs were extracted every $250~\mathrm{fs}$ from the equilibrated trajectory and used for all four electrostatic methods.
For the charge-current analysis, we performed an additional set of ten independent $200~\mathrm{ps}$ microcanonical trajectories using \ac{p3m}, Ewald summation, and \ac{p3maze}. These calculations were designed to resolve the short-time molecular charge response and examine its long-time limit. The microscopic charge current was recorded every five \ac{md} steps, corresponding to a sampling interval of $2.5~\mathrm{fs}$, and the finite-time integrated response was evaluated up to $100~\mathrm{ps}$. 
\ac{pmaze} was not included in this additional long-time charge-current analysis because its \ac{spc} results exhibit a substantially stronger sensitivity to spatial resolution, as independently demonstrated by the diffusion and rotational observables. The purpose of these additional simulations is to establish the expected long-time limit of the charge response and verify that \ac{p3maze} agrees with the established \ac{p3m} and Ewald references. We note that, while the number of molecules and the real-space cutoff were chosen consistently with Ref.~\cite{wu:2006}, those simulations were performed in the (N,P,T) ensemble, whereas the present production calculations are carried out at fixed volume in (N,V,E); differences with the reported reference values may therefore arise from the different thermodynamic protocols.

For the LAMMPS calculations, the real-space cutoff was $9.0~\text{\AA}$, while the Ewald splitting parameter was determined automatically from the requested force accuracy of $10^{-7}$, consistent with the original \ac{spc} parametrization~\cite{wu:2006}. The corresponding splitting parameter was then used to define the Gaussian width of the \ac{p3maze} decomposition ($\sigma=1.87~\text{\AA}$). The real-space cutoff for \ac{p3maze} was selected using
the same convergence procedure described for molten NaCl
in Section 3 of the Supporting Information, yielding
$R_c=6.0$~\AA. A mesh of $40^3$ grid points was found sufficient to converge all structural and dynamical observables discussed below. For \ac{pmaze}, the electrostatic potential was represented on a uniform $180^3$ mesh.

\subsubsection{Observables}
As mentioned above, the \ac{spc} benchmark assesses whether \ac{p3maze} reproduces the structural and dynamical properties obtained with established particle--mesh electrostatic methods. We therefore compare time-resolved observables against the LAMMPS \ac{p3m} implementation and summarize the corresponding integrated quantities for all four electrostatic methods (\ac{p3m}, \ac{p3maze}, Ewald, and \ac{pmaze}) in Table~\ref{tab:water_spcfw_observables}.

\begin{figure}[h!]
    \centering
    \includegraphics[width=0.65\linewidth]{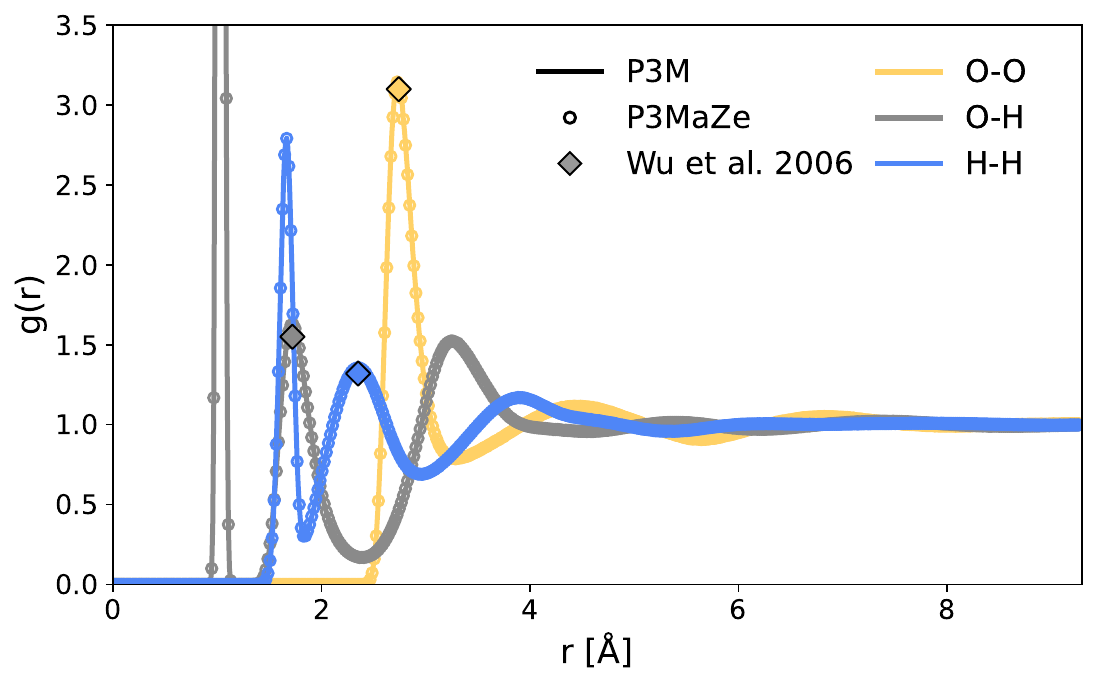}
    \caption{Comparison of oxygen--oxygen, oxygen--hydrogen and hydrogen--hydrogen radial distribution functions for \ac{spc} water obtained with \ac{p3m} and \ac{p3maze}. Diamond symbols indicate the reference peak positions reported by Wu \textit{et al.}~\cite{wu:2006}. The agreement between the curves demonstrates that \ac{p3maze} preserves the equilibrium structure of liquid water.}
    \label{fig:gr_water}
\end{figure}

We first examine the equilibrium liquid structure through the oxygen--oxygen, oxygen--hydrogen, and hydrogen--hydrogen radial distribution functions. Fig.~\ref{fig:gr_water} compares the \ac{p3m} and \ac{p3maze} results together with the reference peak positions reported by Wu \textit{et al.}~\cite{wu:2006}. The two methods are statistically indistinguishable over the full range of intra- and intermolecular separations, demonstrating that the \ac{p3maze} particle--mesh decomposition preserves the equilibrium structure of liquid water.

\begin{figure}[h!]
    \centering
    \includegraphics[width=0.99\linewidth]{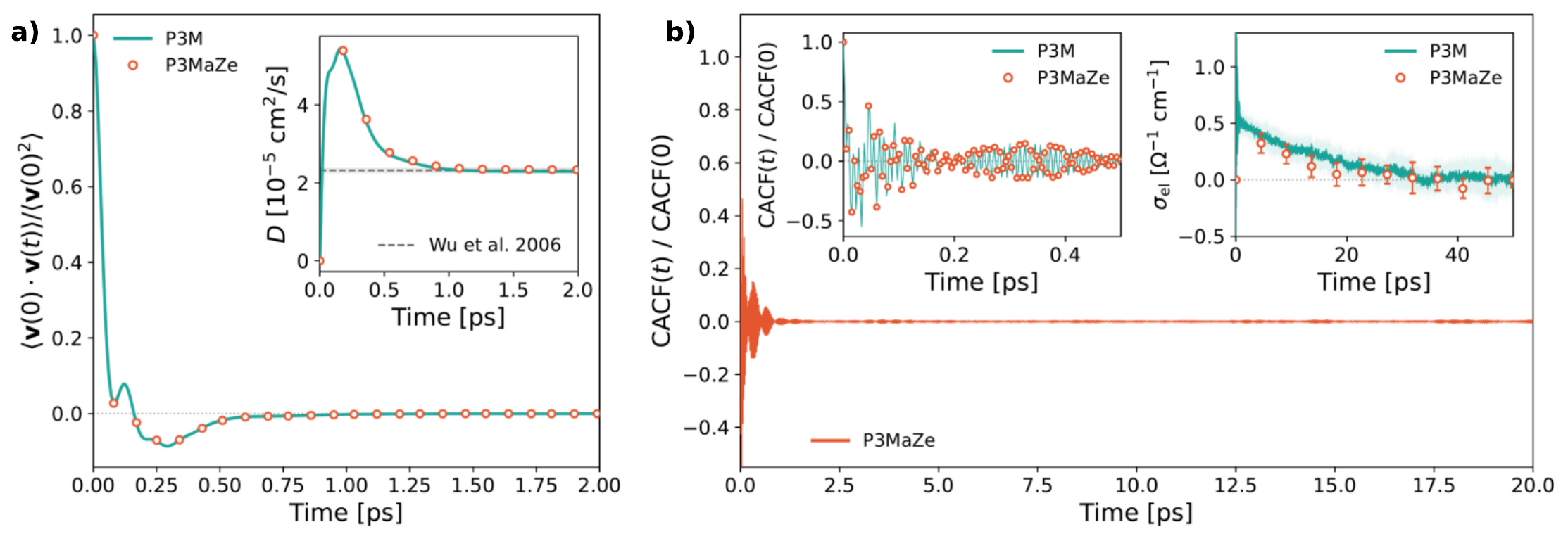}
    \caption{Comparison of single-particle and collective dynamical observables for \ac{spc} water obtained with \ac{p3m} and \ac{p3maze}. \textbf{a)} Normalized molecular \ac{vacf} together with the cumulative Green--Kubo integrals used to determine the self-diffusion coefficient. \textbf{b)} Normalized \ac{cacf}; the main panel shows the first $20$ ps, while the left inset extends the correlation function to $50$ ps and the right inset reports the finite-time integrated charge-current response. \ac{p3maze} results are shown as symbols with error bars, whereas \ac{p3m} is represented by a solid line with a shaded region indicating the standard error of the mean across the independent replicas. The two methods produce statistically indistinguishable correlation functions and transport coefficients, demonstrating that both preserve both translational dynamics and collective electrostatic response.}
    \label{fig:vacf_caf_water}
\end{figure}

As for the dynamical observables, we first consider the molecular \ac{vacf} and the corresponding cumulative Green--Kubo integral for the self-diffusion coefficient. Fig.~\ref{fig:vacf_caf_water} \textbf{a)} shows that the normalized \ac{vacf}s obtained with \ac{p3m} and \ac{p3maze} overlap over the entire correlation time, yielding nearly identical running diffusion coefficients and statistically indistinguishable plateau values. The corresponding numerical estimates are reported in Table~\ref{tab:water_spcfw_observables}, confirming that \ac{p3maze} reproduces the translational dynamics of liquid water with the same accuracy as the standard \ac{p3m} reference.

\begin{figure}[h!]
    \centering
    \includegraphics[width=0.65\linewidth]{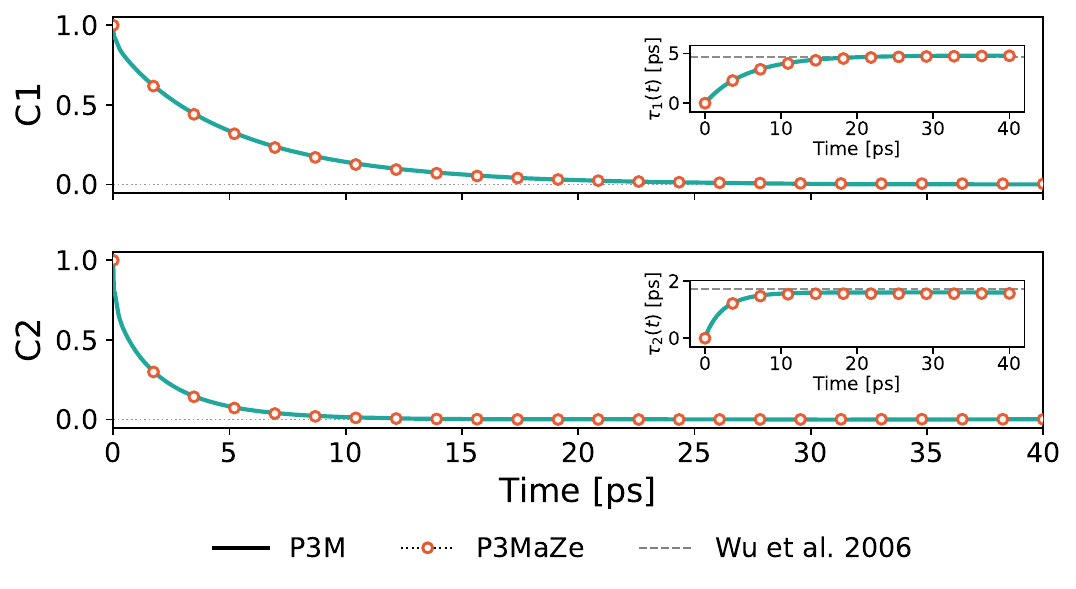}
    \caption{Comparison of dipole rotational relaxation in \ac{spc} water obtained with \ac{p3m} and \ac{p3maze}. The main panel reports the normalized first- and second-order rotational correlation functions, while the inset shows the corresponding cumulative relaxation times. Solid lines denote \ac{p3m} and symbols denote \ac{p3maze}. The close agreement between the two methods demonstrates that \ac{p3maze} accurately reproduces the orientational dynamics of liquid water.}
    \label{fig:water_rotational_relaxation}
\end{figure}

We next examine the collective molecular charge response through the \ac{cacf} and its time integral, shown in Fig.~\ref{fig:vacf_caf_water}\textbf{b)}. Due to the molecular vibrations and librations, the \ac{cacf} exhibits a nonzero oscillatory signal at short times and progressively decays toward zero, as highlighted in the left inset. The corresponding running time integral is shown in the right inset. \ac{p3maze} and  \ac{p3m} results for these quantities agree within statistical uncertainty. For a precise calculation of the residual long-time contribution, we average the running integral of the current autocorrelation function separately for each replica over the interval $80$--$100~\mathrm{ps}$ and estimate the standard error across the ten replica averages. The values obtained with \ac{p3m}, Ewald, and \ac{p3maze}, reported in Table~\ref{tab:water_spcfw_observables}, are all statistically compatible with zero, as expected for a neutral system with no net charge transport. 
These results provide a sensitive probe of the long-range electrostatic correlations, in the same spirit as previous assessments of Coulomb solvers~\cite{Beckers:1998}.
As a final dynamical validation, we consider rotational relaxation. Rotational correlation functions probe molecular reorientation and therefore complement the translational and collective transport observables discussed above. Fig.~\ref{fig:water_rotational_relaxation} reports the first- and second-order rotational correlation functions of the molecular dipole together with their cumulative relaxation times. The corresponding definitions are
\begin{equation}
C_1(t) = \left\langle P_1\left[\mathbf{u}(t)\cdot\mathbf{u}(0)\right]\right\rangle
       = \left\langle \mathbf{u}(t)\cdot\mathbf{u}(0)\right\rangle
\label{eq:rotational_correlation_c1}
\end{equation}
and
\begin{equation}
C_2(t) = \left\langle P_2\left[\mathbf{u}(t)\cdot\mathbf{u}(0)\right]\right\rangle
       = \left\langle \frac{1}{2}\left(3\left[\mathbf{u}(t)\cdot\mathbf{u}(0)\right]^2 - 1\right)\right\rangle,
       \label{eq:rotational_correlation_c2}
\end{equation}
where $\mathbf{u}$ is the molecular unit vector associated with the chosen axis. For the results reported below, $\bm{u}$ is chosen along the molecular dipole direction.
The main panel reports the correlation functions, while the inset shows the cumulative time integrals whose plateau values define the relaxation times $\tau_1$ and $\tau_2$. Excellent agreement between \ac{p3m} and \ac{p3maze} is observed for both the correlation functions and the integrated relaxation times. The corresponding results for the H--H and O--H molecular axes are reported in Section 5 of the Supporting Information, while all integrated relaxation times are summarized in Table~\ref{tab:water_spcfw_observables}.

The integrated observables for all four methods are collected in Table~\ref{tab:water_spcfw_observables}. \ac{p3m}, \ac{p3maze}, and Ewald yield mutually consistent estimates across all quantities, in 
satisfactory agreement with the reference values of Wu et al.~\cite{wu:2006}. \ac{pmaze}, by contrast, systematically overestimates the self-diffusion coefficient and shows larger deviations in the rotational relaxation times, even at the high spatial resolution of $N = 180$ grid points per direction ($h = 0.103~\text{\AA}$). This shows the sensitivity of \ac{pmaze} to grid resolution in systems with weak ionic screening: unlike molten salts, liquid water presents a more demanding long-range electrostatic environment in which the full Coulomb grid problem converges more slowly with mesh refinement. 
\begin{table}[t]
\centering
\caption{Integrated structural and dynamical observables for \ac{spc} water obtained with \ac{p3m}, \ac{p3maze}, Ewald, and \ac{pmaze} at $298.15$ K. \ac{p3maze} is statistically consistent with the established particle--mesh methods across all observables, while the comparison with \ac{pmaze} highlights the improvement obtained by introducing the particle--mesh decomposition. Reference values are taken from Ref.~\cite{wu:2006}, unless stated otherwise.}
\label{tab:water_spcfw_observables}
\footnotesize
\setlength{\tabcolsep}{5pt}
\renewcommand{\arraystretch}{1.12}
\begin{tabular}{llllll}
\hline\hline
Observable & \ac{p3m} & \ac{p3maze} & Ewald & \ac{pmaze} & Ref. value \\
\hline
$D$ [$10^{-5}$ cm$^2$/s]
& $2.29 \pm 0.04$
& $2.33 \pm 0.04$
& $2.30 \pm 0.04$
& $2.56 \pm 0.05$
& $2.32 \pm 0.05$ \\

$\overline{\sigma}$ ($80\text{--}100\,\mathrm{ps})$ [$\Omega^{-1}$/cm]
& $-0.022 \pm 0.031$
& $0.014 \pm 0.023$
& $-0.018 \pm 0.022$
& -- 
& -- \\
\hline
$\tau_1$ (dipole) [ps]
& $4.88 \pm 0.16$
& $4.83 \pm 0.14$
& $4.88 \pm 0.10$
& $4.58 \pm 0.10$
& $4.70$ \\

$\tau_2$ (dipole) [ps]
& $1.60 \pm 0.04$
& $1.58 \pm 0.04$
& $1.62 \pm 0.04$
& $1.51 \pm 0.04$
& $1.72$ \\

$\tau_1$ (H--H) [ps]
& $4.70 \pm 0.12$
& $4.44 \pm 0.10$
& $4.49 \pm 0.11$
& $4.28 \pm 0.10$
& $3.90$ \\

$\tau_2$ (H--H) [ps]
& $2.00 \pm 0.04$
& $1.98 \pm 0.05$
& $2.05 \pm 0.05$
& $1.95 \pm 0.05$
& $2.01$ \\

$\tau_1$ (O--H) [ps]
& $4.79 \pm 0.12$
& $4.60 \pm 0.12$
& $4.52 \pm 0.10$
& $4.38 \pm 0.10$
& $4.17$ \\

$\tau_2$ (O--H) [ps]
& $1.80 \pm 0.04$
& $1.81 \pm 0.04$
& $1.81 \pm 0.04$
& $1.75 \pm 0.04$
& $1.86$ \\
\hline\hline
\end{tabular}
\end{table}

Taken together, these results indicate that \ac{p3maze} preserves not only the structural properties of \ac{spc} water, but also both its single-particle and collective dynamical behavior.

\subsection{Performance and scaling}
\label{subsec:performance}

The objective of this part of the paper is to quantify the computational consequences of the \ac{p3maze} formulation and, in particular, the role of the initial guess generated by the auxiliary-field propagation. We compare \ac{p3maze} with both \ac{pmaze} and an in-house real-space \ac{p3m} implementation. The latter serves as a numerical illustration of the equivalence of the two methods and of their numerical solution provided that a coherent initialization is adopted for the solver. A direct pointwise comparison of the corresponding electrostatic potentials, including their dependence on rigid translations relative to the mesh, is reported in Section 6 of the Supporting Information to further substantiate this equivalence. 

\begin{figure}[h!]
\centering
\includegraphics[width=0.95\linewidth]{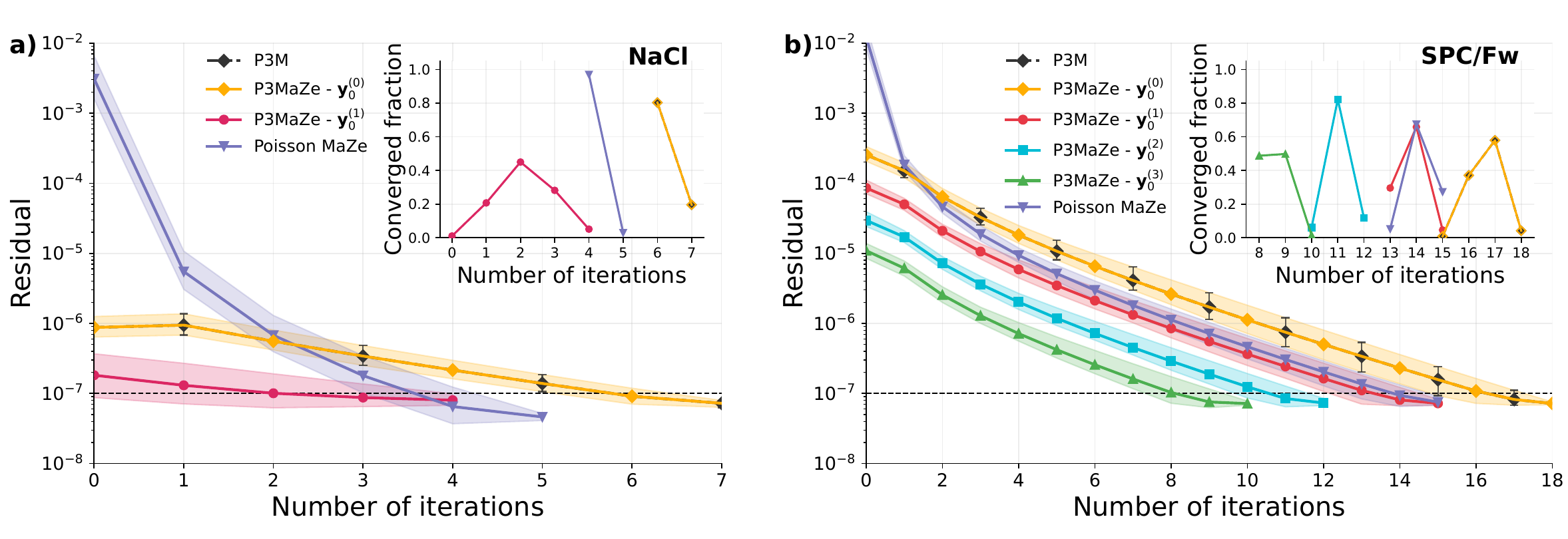}
\caption{Residual convergence of the long-range electrostatic solver for
\textbf{a)} molten NaCl and \textbf{b)} \ac{spc} water.
Lines denote the mean residual decay. The minimum-to-maximum spread over
sampled \ac{md} steps and restart configurations is represented by shaded
regions for \ac{p3maze} and \ac{pmaze}, and by error bars for \ac{p3m}.
With the zero-correction predictor, \ac{p3maze} reproduces the convergence of the corresponding real-space \ac{p3m} solver, whereas predictors constructed from the correction history progressively reduce the number of iterations. The inset in each panel reports the iteration-count distribution over the sampled data.}
\label{fig:iteration_moltensalt}
\end{figure}

In Fig.~\ref{fig:iteration_moltensalt} we show the average maximum-norm residual as a function of the number of iterations for NaCl (left panel) and \ac{spc} water (right panel) computed with \ac{pmaze} (purple curves) and \ac{p3maze} (different colors identify different initializers for the iterative multigrid solver). The maximum-norm residual is defined, for all methods, as: 
\begin{equation} 
\mathrm{res}_{\nu} = \max_n \left| \left\{ \left( \frac{h}{4\pi}\bm M \bm\phi_{\rm LR}^{\nu} + \bm q^{s} \right)_n \right\} \right|. 
\label{eq:solver_residual}
\end{equation}
where $\nu$ is the index of the iteration and $n$ runs over the mesh. 

\begin{figure}[h!] 
    \centering \includegraphics[width=0.95\linewidth]{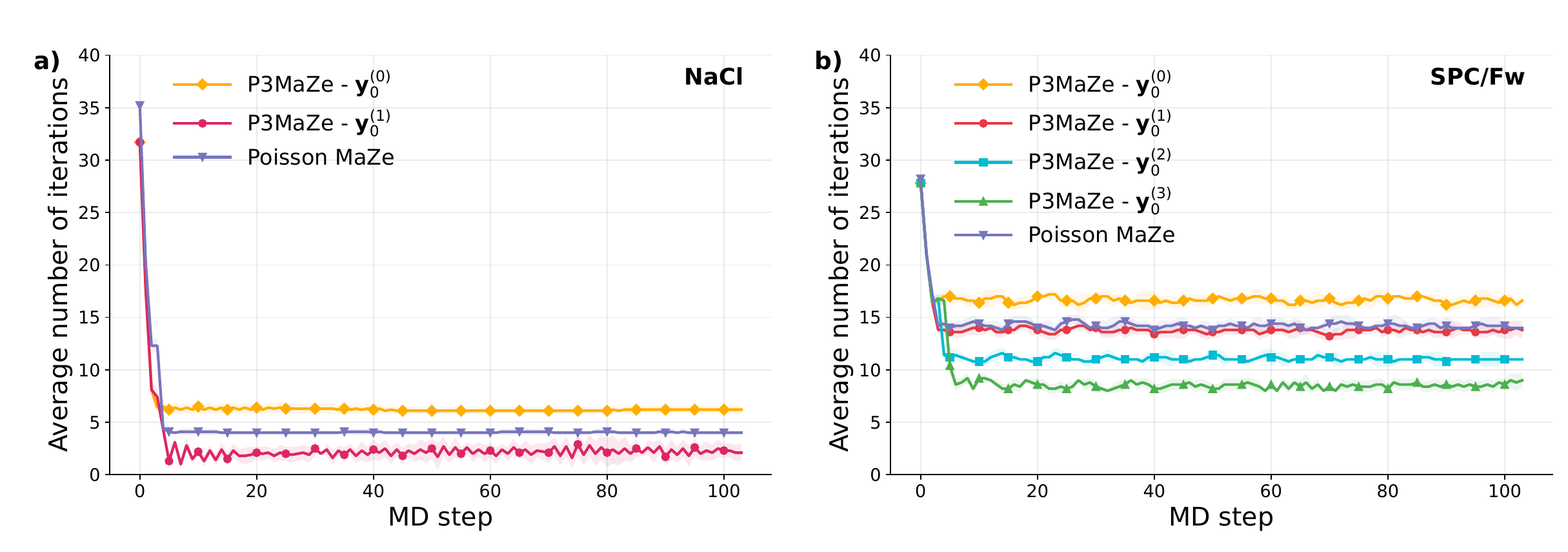} 
    \caption{Average number of multigrid iterations per \ac{md} step for \textbf{a)} molten NaCl and \textbf{b)} \ac{spc} water. The faster residual convergence observed in Fig.~\ref{fig:iteration_moltensalt} translates into a sustained reduction in the number of multigrid iterations during \ac{md}, and higher-order predictors progressively improve the \ac{p3maze} convergence in the more demanding \ac{spc} system.} \label{fig:predictor_moltensalt_water} 
\end{figure}

The data were obtained from the same restart configurations used for the independent dynamical replicas discussed in the previous subsections. For each restart and each method, short trajectories of 100 \ac{md} steps were propagated, and the residual was recorded after each multigrid iteration during the field update. The initial steps of each trajectory were discarded because they are affected by the initialization of the field solver and converge more slowly than the subsequent plateau regime (see also Fig.~\ref{fig:predictor_moltensalt_water}). The plotted curves therefore report the residual decay averaged over the remaining steps and restart configurations. The minimum-to-maximum spread across the analyzed samples is shown by shaded regions for \ac{p3maze} and \ac{pmaze}, and by error bars for \ac{p3m}.

In both panels of Fig.~\ref{fig:iteration_moltensalt}, the yellow curve reports results for \ac{p3maze} initialized with $\bm{y}^{(0)}_0=0$. The black symbols superimposed on this curve are obtained via a direct solution of the discretized Poisson equation with the same multigrid solver initialized according to Eq.~\eqref{eq:potVerletinit}. The excellent agreement of the data illustrates the \ac{p3m}/\ac{p3maze} equivalence. For the molten salt, Fig.~\ref{fig:iteration_moltensalt} \textbf{a)}, 
we also show that, when the value of the SHAKE correction at the previous time step, $\bm{y}^{(1)}_0$, is instead used as the initial guess, \ac{p3maze} reaches the prescribed tolerance in one or two iterations, and always within one to four iterations over the sampled steps. By contrast, \ac{pmaze} requires four to five iterations. These results are more clearly visible in the inset in the figure that reports the iteration count distribution for the sampled data. This acceleration reflects the combined effect of the \ac{maze} correction mechanism and the smoother long-range source term entering the \ac{p3maze} constraint. \\

The \ac{spc} system, Fig.~\ref{fig:iteration_moltensalt} \textbf{b)}, provides a more demanding test of the iterative solver. All methods require more iterations than in molten NaCl, reflecting the weaker electrostatic screening of liquid water and the larger integration time step, both of which reduce the correlation between successive mesh updates. As in molten NaCl, \ac{p3maze} with $\bm{y}_0^{(0)}=0$ closely reproduces the convergence of the direct \ac{p3m} solve, consistently with the algebraic equivalence discussed above. This system also allows us to better assess different initializations for the iterative SHAKE solution. 

In addition to the reference choice $\bm y_0^{(0)}=\bm 0$, we consider the cases $m=1,2,$ and $3$ of the correction predictor defined in Eq.~\eqref{eq:correction_predictor}. As established in Eq.~\eqref{eq:chronological_phi_predictor}, these choices correspond to quadratic, cubic, and quartic chronological extrapolations of the converged mesh potential, respectively. Together with the linear Verlet prediction associated with $\bm y_0^{(0)}$, they provide a systematic hierarchy of increasingly accurate temporal initializations.
Using $\bm y_0^{(1)}$, \ac{p3maze} converges in approximately 13--14 iterations, already below the baseline of $\bm y_0^{(0)}$/\ac{p3m} of 16--17 iterations. \ac{pmaze} requires 17--19 iterations and therefore offers no convergence advantage over \ac{p3m} for this weakly screened molecular system. Higher-order predictors substantially improve the \ac{p3maze} convergence: $\bm y_0^{(2)}$ reduces the iteration count to approximately 11, while $\bm y_0^{(3)}$ further reduces it to 8--9, nearly halving the number of iterations relative to the direct real-space \ac{p3m} solve. The systematic reduction with predictor order demonstrates how the solver acceleration is controlled by the quality of the temporal initialization. 

\begin{figure}[h!]
\centering
\includegraphics[width=0.45\linewidth]{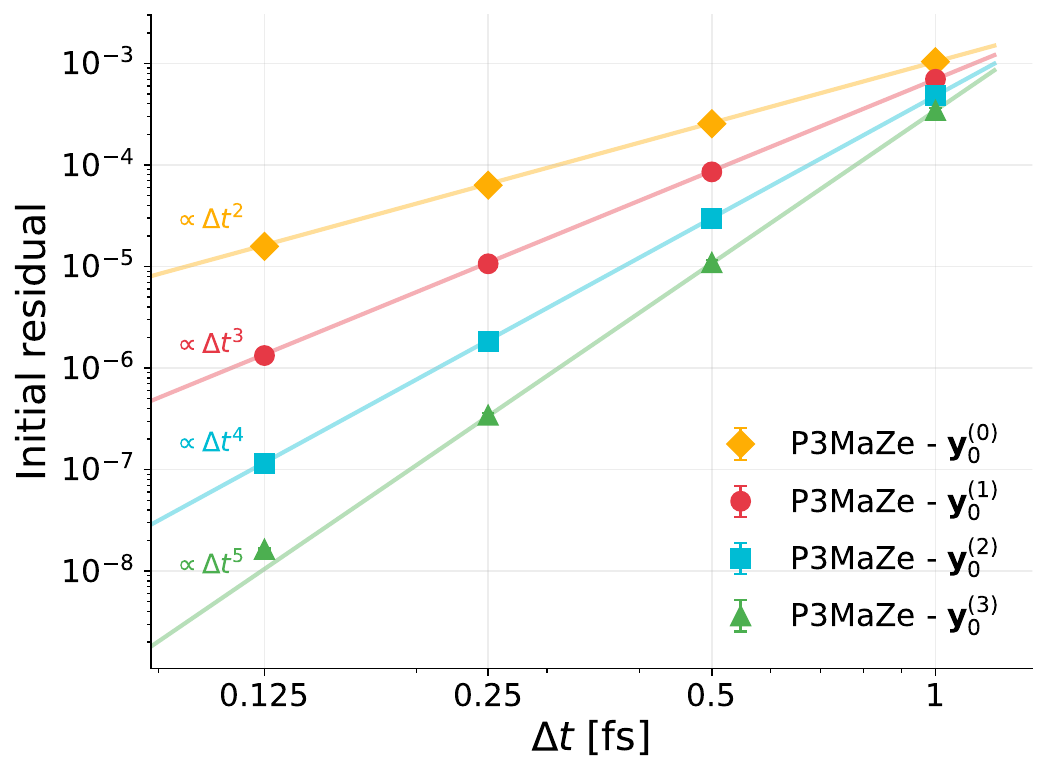}
\caption{Initial residual of the long-range electrostatic solve for \ac{spc} water as a function of the integration time step $\Delta t$. Results are shown for the correction predictors $\bm y_0^{(m)}$, with $m=0,1,2,3$. The reference lines indicate the expected asymptotic scalings $\Delta t^{m+2}$.}
\label{fig:sigma_p_scaling}
\end{figure}

This dependence can be further examined considering the behavior of the initial residue as a function of the time step $\Delta t$ employed in the MD trajectory. For any time-dependent quantity with a bounded $(m+2)$th derivative over the extrapolation interval, the standard error formula for polynomial interpolation predicts that a degree-$(m+1)$ Lagrange extrapolation has an error of order $\Delta t^{m+2}$; see Ref.~\cite{stoer:1980}, Theorem~2.1.4.1. Accordingly, the residual is expected to decrease with increasing predictor order according to the corresponding power-law scaling in $\Delta t$. Figure~\ref{fig:sigma_p_scaling} shows that the theoretical trend is very closely matched by our simulations. (Data reported for the case of \ac{spc} water.) The result can be attributed to the smooth variation of the potential with time driven in turn by the smooth variations of the mesh charge that follow the continuous changes in the ions positions along the trajectory and is preserved by our mesh assignment and smoothing operations.

\begin{figure}[t!]
    \centering    \includegraphics[width=0.99\linewidth]{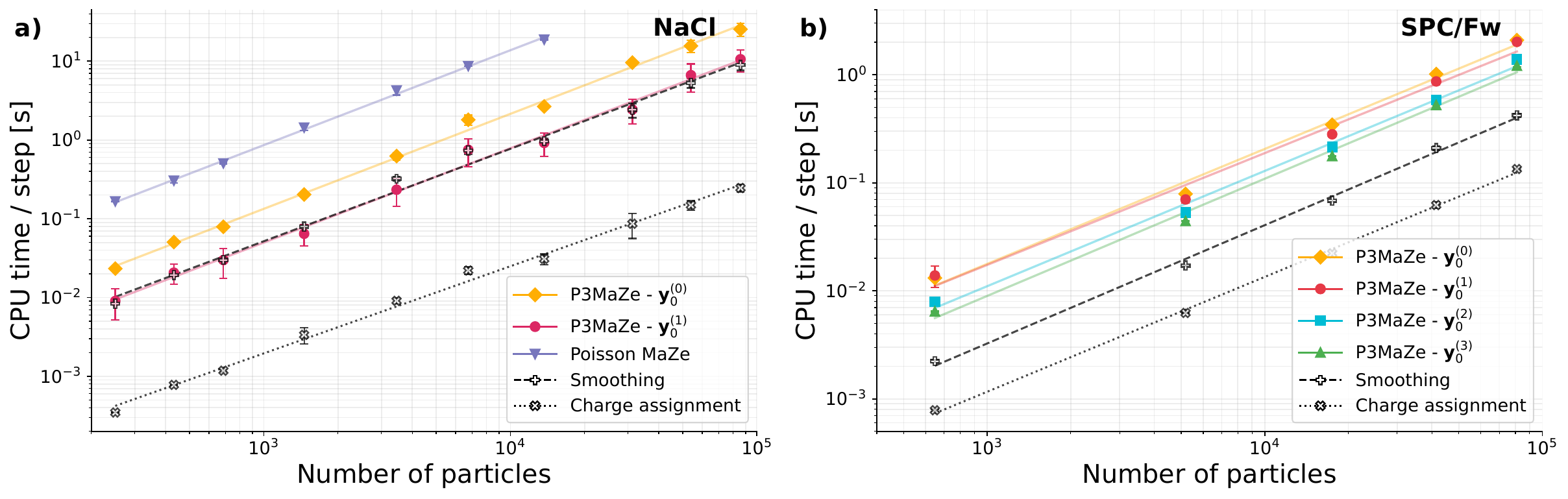}
    \caption{Electrostatic CPU time per \ac{md} step as a function of system size shown on logarithmic axes. Linear fits confirm the expected $\mathcal{O}(N)$ scaling of all
real-space formulations, while the vertical separation between the curves
quantifies the reduction in computational cost obtained with
\ac{p3maze}. Shared charge-assignment and smoothing costs are shown
separately.
\textbf{a)} Molten NaCl.
\textbf{b)} \ac{spc} water.}
    \label{fig:performance_scaling}
\end{figure}

The reduction in the initial residual shown in Fig.~\ref{fig:sigma_p_scaling} translates into a stable, lower number of multigrid iterations during \ac{md}, as shown in Fig.~\ref{fig:predictor_moltensalt_water}. In both systems, the first steps exhibit a short initialization transient before reaching a stable plateau. Within this regime, the iteration count decreases systematically with predictor order, with the largest reduction observed for \ac{spc} water.
\\
We conclude our analysis by demonstrating the linear scaling of the algorithm with the overall number of degrees of freedom. Fig.~\ref{fig:performance_scaling} reports the electrostatic CPU time per \ac{md} step as a function of system size on logarithmic axes for molten NaCl
(panel~a) and \ac{spc} water (panel~b). The mesh resolution was scaled to preserve the mesh spacing $h$ of the corresponding production simulations as $N_p$ was increased. For each system, all simulations were performed at the same density. Linear fits in the log--log representation confirm the expected $\mathcal{O}(N)$ scaling of the real-space formulation, while the vertical offsets between the curves quantify the reduction in computational cost achieved by the different initializations. We note that, in addition to the properties of the multigrid solver, linear scaling is obtained due to the sparse nature of the discretized Laplacian with PBC that only has the same seven non zero entries per row~\cite{note:sparsity}.
% \bibnote{The sparsity of the matrix and its $\mathcal{O}(N)$ is preserved in different boundary conditions.}

This enables to perform matrix vector multiplication with $\mathcal{O}(N)$ scaling and very low memory requirements. In the present implementation, $\bm M$ is not explicitly assembled or stored. Its action is evaluated directly through the seven-point finite-difference stencil applied to the vector. Furthermore, the storage required by the different initialization of the iterative cycle is respectively one, two, and three previous $\bm y$ vectors, each containing $\mathcal{N}$ elements; thus, the predictor does not alter the $O(\mathcal{N})$ memory requirements.

For molten NaCl, Fig.~\ref{fig:performance_scaling} \textbf{a)}, all methods exhibit approximately unit slopes, confirming the expected linear scaling of the underlying multigrid solver. \ac{pmaze} is the most expensive method. Considering the mesh-solver contribution alone, linear fits of the untransformed CPU times as functions of system size yield a timing coefficient approximately $19$ times larger than that of \ac{p3maze}, reflecting the cost of solving the full Coulomb problem on the fine $120^3$ reference grid. When charge assignment and smoothing are included, the ratio between the total electrostatic timing coefficients decreases to approximately $9$, since smoothing accounts for about half of the total \ac{p3maze} cost. The corresponding solver-only and total electrostatic fit coefficients are reported in Table S3 of the Supporting Information.
Identical results would be found for a real-space \ac{p3m} adopting the chronological initialization of the potential. As the improved initialization reduces the long-range solver cost, the smoothing step becomes the dominant contribution to the \ac{p3maze} electrostatic time, accounting for approximately 50\% of the total cost across the system sizes considered.
The same linear scaling is observed for \ac{spc} water, Fig.~\ref{fig:performance_scaling} \textbf{b)}. Here, \ac{pmaze} is omitted because the fine mesh required for the full Coulomb grid problem makes it considerably expensive at the larger system sizes.  For this system, smoothing  rises to 15--30\% of the computational overhead in the \ac{p3maze} variants as the solver contribution decreases with improved initial guess.

These results show that the particle--mesh decomposition and the \ac{maze} propagation play complementary roles. The decomposition reduces the size and smooths the source of the long-range mesh problem, while the temporal information naturally available in the \ac{maze} dynamics provides increasingly accurate initial guesses for the iterative solve. The resulting method preserves the systematic accuracy controls of particle--mesh electrostatics while reducing the cost of the long-range solver. The timing analysis also identifies the next optimization target: once the multigrid correction is accelerated, particle--mesh preprocessing, and in particular the smoothing step, becomes the main remaining bottleneck.

\section{Conclusions}

We have introduced \ac{p3maze}, a real-space particle--mesh electrostatic method that combines the standard short-range/long-range decomposition of \ac{p3m} with the Mass-Zero constrained-dynamics framework. In this formulation, the smooth long-range electrostatic potential is treated as a zero-inertia auxiliary field whose evolution is constrained by the discretized Poisson equation. As a result, the long-range mesh problem is solved through the constrained evolution of the auxiliary variables. This formulation combines the complementary strengths of particle--mesh electrostatics and constrained dynamics. By retaining the standard \ac{p3m} decomposition, \ac{p3maze} preserves the systematic accuracy controls associated with the real-space cutoff, the Ewald splitting, the mesh spacing, and the charge-assignment procedure. At the same time, casting the long-range solve as a holonomic constraint on zero-inertia auxiliary variables endows \ac{p3maze} with the structural properties of constrained \ac{maze} dynamics: the equations of motion are symplectic and time-reversible, the auxiliary field remains adiabatically separated from the physical degrees of freedom in the $\mu\rightarrow0$ limit, and the resulting trajectories exhibit the expected long-term stability. Within this structurally robust setting, the auxiliary-field propagation additionally provides effective initial guesses for the iterative long-range solve.

The numerical benchmarks demonstrate that this reformulation efficiently preserves the physical accuracy of particle--mesh electrostatics. For molten NaCl, \ac{p3maze} reproduces the structural and dynamical properties obtained with the original \ac{pmaze} formulation and with literature reference data. For flexible \ac{spc} water, a substantially more demanding electrostatic benchmark, \ac{p3maze} yields structural, translational, collective, and rotational dynamical observables that are statistically indistinguishable from those obtained with LAMMPS \ac{p3m} and Ewald summation.

The solver analysis clarifies the origin of the low iterative cost observed in \ac{p3maze}, and in doing so establishes a new result for the \ac{maze} framework itself. For the linear, time-independent Poisson operator considered here, we show that the \ac{maze} correction problem is algebraically equivalent to the corresponding direct Poisson solve, with the two connected by an explicit, closed-form chronological extrapolation of the mesh potential. This analysis identifies, for the first time, the precise mechanism responsible for the accelerated convergence previously observed in \ac{maze}-based electrostatic solvers, turning what had been an empirical feature of the formalism into a controlled and predictable one. Because the extrapolation formula is independent of the constrained-dynamics construction itself, our results indicate that \ac{maze} dynamics identifies effective initializations for conventional iterative solvers. In both benchmark systems, exploiting this temporal information reduces the number of multigrid iterations substantially, and higher-order predictors constructed along the same lines provide additional gains in the more demanding \ac{spc} simulations. The timing analysis confirms that \ac{p3maze} preserves the expected linear scaling of real-space particle--mesh electrostatics while reducing the prefactor associated with the long-range mesh solve. As the cost of the constrained long-range solve decreases, charge assignment and smoothing become an increasingly important fraction of the total electrostatic cost, identifying the particle--mesh preprocessing stage as the next target for algorithmic optimization.

The results presented in this paper provide a flexible foundation for further algorithmic developments for \ac{p3maze} and equivalent \ac{p3m} solvers, including improved particle--mesh preprocessing, more effective predictor strategies, and implementations targeting highly parallel many-core and GPU architectures.

\section*{Acknowledgements}
The authors thank Benoît Roux for seeding the idea of using \ac{maze} for electrostatic calculations. We are also grateful to the reviewers of this work: their insightful comments led to deeper understanding of the convergence properties of \ac{p3maze} and its relationship with conventional \ac{p3m} solvers.

This research was supported by the NCCR MARVEL, a National Centre for Competence in Research, funded by the Swiss National Science Foundation (grant number 205602). 

\section*{Supporting information}
The Supporting Information contains the derivation of the \ac{p3maze} equations of motion and numerical integration scheme; NVE stability and conservation tests; convergence tests and parameter tuning for \ac{p3maze}; \ac{spc} force-field parameters; additional rotational relaxation functions for the H--H and O--H molecular axes; a direct pointwise comparison of the \ac{p3m} and \ac{p3maze} electrostatic potentials, including their translational dependence relative to the mesh; and numerical details of the performance scaling analysis.

\section*{Data and Software Availability}

The data and software underlying this study are available through Materials Cloud~\cite{talirz2020materials} at \url{https://archive.materialscloud.org/records/x415p-e7q15}. The archive includes the source code, simulation input files, processed data, and analysis scripts required to reproduce the results reported in the manuscript and Supporting Information.
\printbibliography
\newpage

% \rule{0.05in}{1.75in}%
\begin{minipage}[b][1.75in]{3.25in}

  \centering
  \includegraphics[width=3.25in]{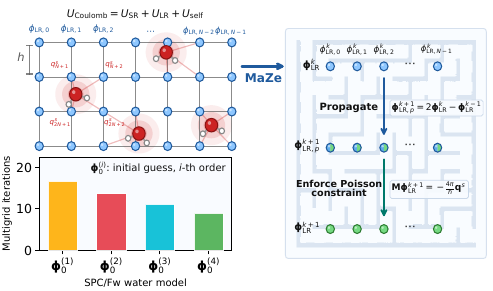}
  \par\small For Table of Contents Only
\end{minipage}%
% \rule{0.05in}{1.75in}

\end{document}

\maketitle
\section{\ac{p3maze} evolution equations}
\label{si:eq_motion}

We derive the Mass Zero equations of motion from the extended Lagrangian of Eq.~\eqref{main-eq:p3maze_lagrangian} of the main text,
\begin{equation}
\begin{split}
    \mathcal{L}_{\rm \ac{p3maze}}
    = &\frac{1}{2} \sum_{\beta=1}^{N_p} m_\beta \dot{\bm{r}}_\beta^T \dot{\bm{r}}_\beta + \frac{1}{2} \mu \dot{\bm\phi}_{\rm LR}^T\dot{\bm\phi}_{\rm LR}
    \\
    &-\bar{U}(\{\bm{r}_\beta\})
    - U_{\rm SR}(\{\bm{r}_\beta\})
    - U_{\rm LR}(\bm\phi_{\rm LR};\{\bm{r}_\beta\})
    - U_{\rm self}
    - \bm\lambda^T \bm\sigma_{\rm LR}(\bm\phi_{\rm LR};\{\bm{r}_\beta\}),
\end{split}
\end{equation}
where $\mu$ is the fictitious inertia of the auxiliary degrees of freedom $\bm\phi_{\rm LR}$, $\bm\lambda$ is the vector of Lagrange multipliers enforcing the constraint $\bm\sigma_{\rm LR}=\bm 0$ [Eq.~\eqref{main-eq:p3maze_constraint} of the main text], $U_{\rm LR}$ is the discretized long-range electrostatic energy [Eq.~\eqref{main-eq:p3maze_vlr_discrete} of the main text], and $U_{\rm self}$ is the constant Gaussian self-energy defined in Eq.~\eqref{main-eq:p3maze_vself_general} of the main text.
The Euler--Lagrange equations are
\begin{equation}
\begin{split}
    m_\alpha \ddot{\bm{r}}_\alpha
    & =
    -\nabla_{\bm{r}_\alpha} U_{\rm SR}(\{\bm{r}_\beta\})
    -\nabla_{\bm{r}_\alpha} U_{\rm LR}(\bm\phi_{\rm LR};\{\bm{r}_\beta\})
    -\nabla_{\bm{r}_\alpha} \bar{U}(\{\bm{r}_\beta\})\\&
    -\sum_n \lambda_n
    \nabla_{\bm{r}_\alpha} \sigma_{{\rm LR},n}(\bm\phi_{\rm LR};\{\bm{r}_\beta\}), 
    \\
     \mu\,\ddot\phi_{{\rm LR},n}
    &=
    -\frac{\partial U_{\rm LR}}{\partial \phi_{{\rm LR},n}}
    -\sum_m \lambda_m
    \frac{\partial \sigma_{{\rm LR},m}}{\partial \phi_{{\rm LR},n}}.
 \end{split}   \label{eq:si_eom_particles}
\end{equation}
Along the trajectory, the constraints
\begin{equation}
    \sigma_{{\rm LR},n} = \frac{\partial U_{\rm LR}}{\partial \phi_{{\rm LR},n}} = \left(\frac{h}{4\pi}\bm{M}\bm\phi_{\rm LR} + \bm{q}^s\right)_n = 0
\end{equation}
are satisfied so the force term $-\partial U_{\rm LR}/\partial \phi_{{\rm LR},n}$ in the equation for the auxiliary variables is null. We also have $\frac{\partial \sigma_{{\rm LR},m}}{\partial \phi_{{\rm LR},n}}=\frac{h}{4\pi}M_{mn}$. Using the symmetry of the matrix $\bm M$ and dividing by the inertia $\mu$, we obtain for the evolution equation of the auxiliary variables
\begin{equation}
    \ddot{\bm\phi}_{\rm LR}
    =
    -\frac{h}{4\pi}\bm{M}{\frac{\bm\lambda}{\mu}}.
\end{equation}
We now take the limit $\mu \to 0$. For $\ddot{\bm\phi}_{\rm LR}$ to remain finite in this limit, we must have $\bm\lambda \to \bm 0$ as $\mu \to 0$~\cite{coretti:2018b, troni_mass-zero_2025}. We therefore define the rescaled multiplier $\bm\eta = \lim_{\mu \to 0} \frac{h}{4\pi}\bm\lambda/\mu$, which remains finite in the zero--inertia limit for the auxiliary variables. With this definition, we have
\begin{equation}
    \ddot{\bm\phi}_{\rm LR} = -\bm{M}\bm\eta.
\end{equation}
Note that, for $\bm\lambda \to \bm 0$, the constraint force term in
Eq.~\eqref{eq:si_eom_particles} vanishes, and the equations
of motion for the physical degrees of freedom reduce to
\begin{equation}
    m_\alpha \ddot{\bm{r}}_\alpha
    =
    -\nabla_{\bm{r}_\alpha} U_{\rm SR}(\{\bm{r}_\beta\})
    -\nabla_{\bm{r}_\alpha} U_{\rm LR}(\bm\phi_{\rm LR};\{\bm{r}_\beta\})
    -\nabla_{\bm{r}_\alpha} \bar{U}(\{\bm{r}_\beta\}),
\end{equation}
recovering exact physical dynamics and leading to Eq.~\eqref{main-eq:MaZeEvolution} of the main text.

\subsection*{Numerical integration}

The equations of motion above are integrated numerically using
standard molecular dynamics algorithms~\cite{coretti:2018b,
troni_mass-zero_2025}. The physical degrees of freedom
$\bm{r}_\alpha$ are evolved via velocity Verlet (or OVRVO for $(N,V,T)$ simulations), while the
auxiliary variables $\bm\phi_{\rm LR}$ are propagated via the
standard Verlet algorithm,
\begin{equation}
    \bm\phi_{\rm LR}^{k+1}
    =
    2\bm\phi_{\rm LR}^k - \bm\phi_{\rm LR}^{k-1}
    - \Delta t^2\bm{M}\bm\eta,
\end{equation}
where $k$ denotes the time-step index and $\Delta t$ is the
timestep. Identifying the Verlet prediction
$\bm\phi_{{\rm LR},p}^{k+1} = 2\bm\phi_{\rm LR}^k -
\bm\phi_{\rm LR}^{k-1}$ as the provisional value, the
time-updated field is
\begin{equation}
    \bm\phi_{\rm LR}^{k+1}
    =
    \bm\phi_{{\rm LR},p}^{k+1}
    -
    \Delta t^2\bm{M}\bm\eta.
    \label{eq:si_verlet_phi}
\end{equation} 
The rescaled Lagrange multipliers
$\bm\eta$ are determined at each step by enforcing
$\bm\sigma_{\rm LR}(\bm\phi_{\rm LR}^{k+1};\{\bm{r}_\alpha^{k+1}\})
= \bm 0$. Substituting Eq.~\eqref{eq:si_verlet_phi} into the
constraint, we obtain
\begin{equation}
    \frac{h}{4\pi}\bm{M}\bm\phi_{\rm LR}^{k+1} + \bm{q}^{s,k+1}
    = \bm 0 \implies \frac{h}{4\pi}\bm{M}\bm\phi_{{\rm LR},p}^{k+1} + \bm{q}^{s,k+1} = \frac{h}{4\pi}\Delta t^2 \bm{M}^2\bm\eta.
\end{equation}
Defining
$\bm y^{k+1}
=
\frac{h}{4\pi}\Delta t^2
\bm M\bm\eta^{k+1}$,
the equation above reduces to
\begin{equation}
\bm M\bm y^{k+1}=\bm\sigma_{{\rm LR},p}^{k+1},
\end{equation}
which is the linear system of
Eq.~\eqref{main-eq:correction_system} of the main text. Once
$\bm y^{k+1}$ is obtained, the updated field follows as
\begin{equation}
\bm\phi_{\rm LR}^{k+1}
=
\bm\phi_{{\rm LR},p}^{k+1}
-
\frac{4\pi}{h}\bm y^{k+1}.
\end{equation}
This system is solved iteratively at each time step using the multigrid solver described in the main text.
Since the provisional potential is constructed from previously gauge-fixed fields, it also has zero spatial mean. The gauge $\langle\phi_{\rm LR}\rangle=0$ is therefore preserved by selecting the zero-mean solution of the correction equation for $\bm y^{k+1}$. In practice, the spatial mean of $\bm y^{k+1}$ is removed during the correction step, and any residual mean is subtracted from the updated potential to prevent numerical drift. Solvability of the periodic Poisson equation additionally requires the mesh source to have zero spatial average. This condition is satisfied for the charge-neutral systems considered in this work. Systems with nonzero total particle charge can be treated by adding a uniform compensating background to the mesh source before constructing the Poisson constraint, as conventionally done in periodic Ewald and particle--mesh methods.

\section{NVE stability and conservation properties}
\label{sec:nve_stability}
\begin{figure}[H]
    \centering
    \includegraphics[width=0.82\textwidth]{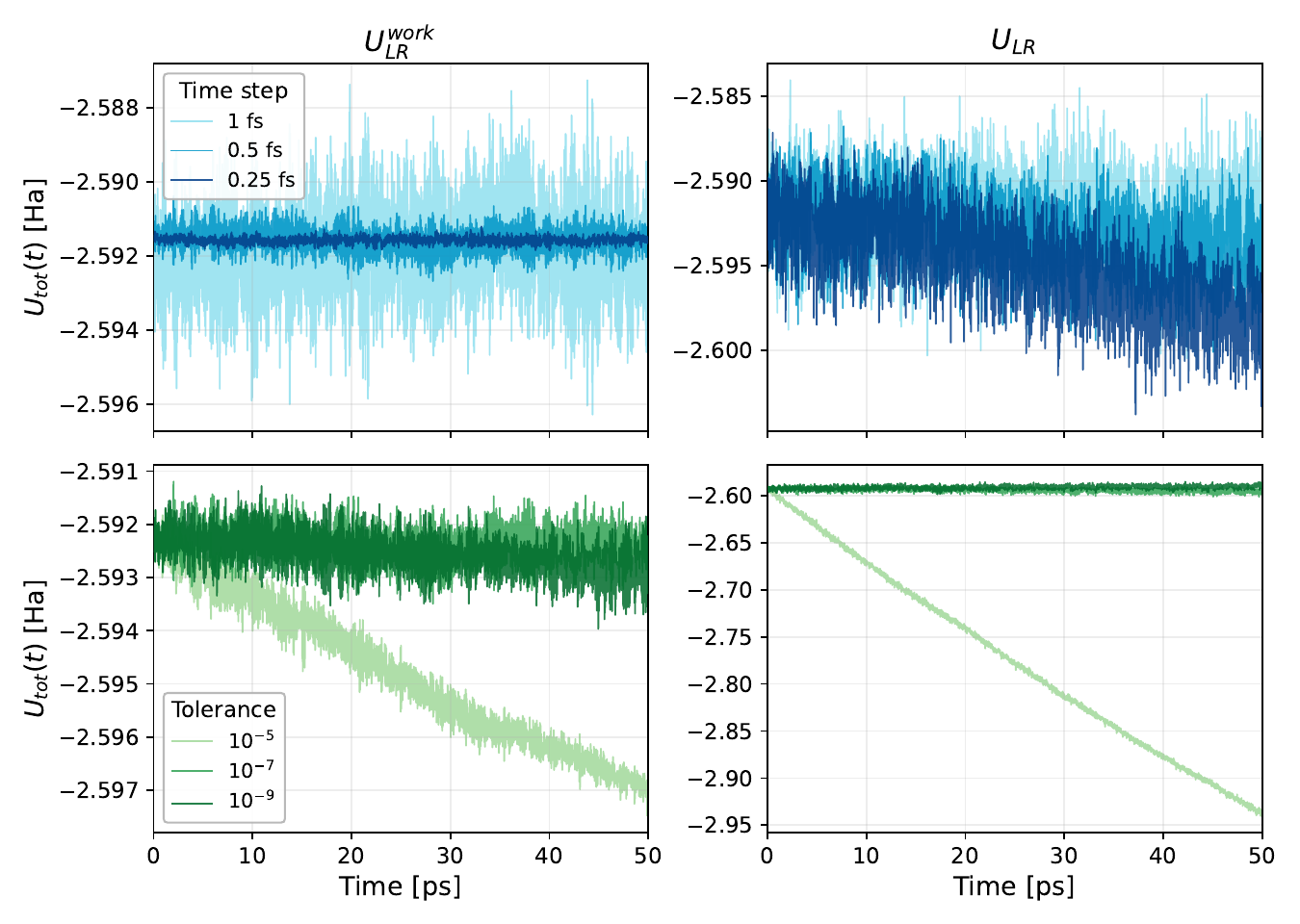}
    \caption{
    Comparison of the work-based and explicit total-energy estimators for SPC/Fw water simulated with \ac{p3maze}. The upper panels show the time-step sweep at fixed constraint tolerance $10^{-7}$, while the lower panels show the tolerance sweep at fixed time step $0.5$~fs. The work-based estimator is consistent with the electrostatic forces employed in the dynamics and exhibits systematic convergence with both numerical parameters.
    }
    \label{fig:water_energy_estimators}
\end{figure}

We assessed the NVE stability of \ac{p3maze} using the SPC/Fw water system described in the main text. The system contains 216 water molecules in a cubic periodic cell of side $18.552$~\AA. All trajectories were started from the same equilibrated configuration at $298.15$~K, propagated for $50$~ps, and sampled every $10$~fs.
For \ac{p3maze}, the total energy was evaluated both using the electrostatic work accumulated from the forces employed in the dynamics (see Eq.~\eqref{main-eq:electrostatic_work} of the main text) and the explicit estimator constructed from the long-range mesh term (Eq.~\eqref{main-eq:p3maze_vlr_on_constraint}). 
Figure~\ref{fig:water_energy_estimators} compares the two quantities.
The work-based energy exhibits the expected improvement as the time step and constraint tolerance are reduced. By contrast, the explicit estimator retains finite-mesh fluctuations that do not decrease systematically with the time step. These fluctuations arise from the finite-grid representation of the mesh energy and its non-exact correspondence with the interpolated force. The analytical Gaussian self-energy is constant for fixed charges and screening width and therefore does not contribute to the observed drift. We therefore use $U^{\mathrm{work}}_{\rm LR}$ for the following NVE conservation analysis. As this estimator is reconstructed directly from the forces, however, its conservation is not an independent test of dynamical stability: by the work--energy theorem, changes in kinetic energy are compensated by opposite changes in the accumulated work, up to numerical integration and work-quadrature errors. A complementary and independent convergence analysis based directly on the kinetic energy, including its dependence on the integration time step and solver tolerance, was reported for \ac{pmaze} in Ref.~\cite{troni_mass-zero_2025}.

\begin{figure}[H]
    \centering
    \includegraphics[width=\textwidth]{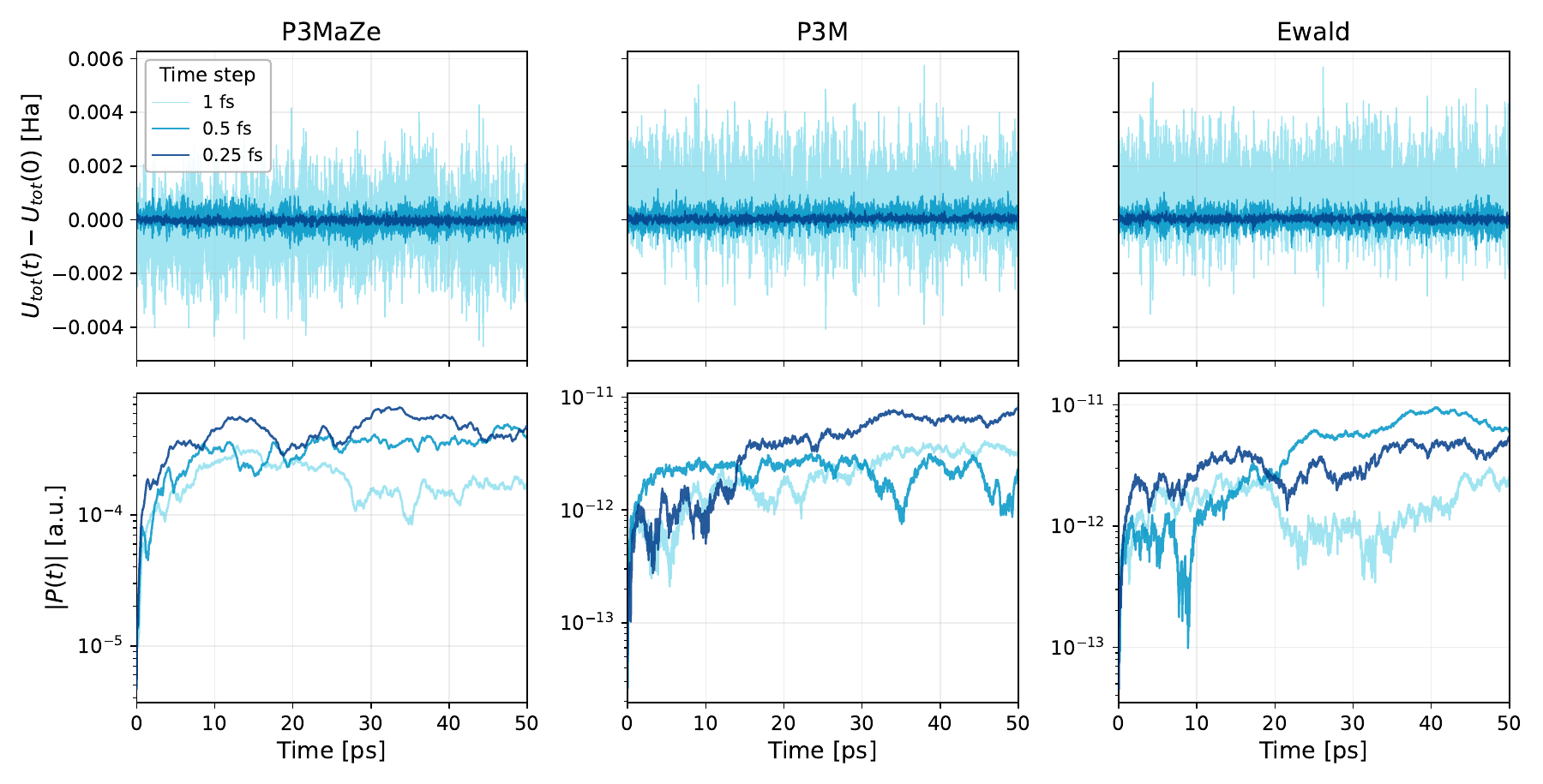}
    \caption{
    NVE stability of SPC/Fw water as a function of the integration time step for \ac{p3maze}, P3M, and Ewald. The upper panels show the total-energy change relative to its initial value, while the lower panels show the norm of the total momentum. All trajectories span $50$~ps. The energy fluctuations decrease systematically with $\Delta t$ and are comparable among the three methods. No periodic momentum removal was applied during the trajectories.
    }
    \label{fig:water_nve_timestep}
\end{figure}

We first examine the dependence of the work-based NVE energy conservation on the integration time step. Simulations were performed with $\Delta t=1.0$, $0.5$, and $0.25$~fs. The \ac{p3maze} constraint tolerance was fixed at $10^{-7}$. Equivalent trajectories were propagated using the P3M and Ewald implementations in LAMMPS, with the long-range accuracy fixed at $10^{-6}$~\cite{note:lammps_accuracy}.
Figure~\ref{fig:water_nve_timestep} reports the total-energy change and total-momentum norm for the three electrostatic methods. For all methods, the energy fluctuations decrease systematically as the time step is reduced. The relative fluctuations of the \ac{p3maze} total energy decrease from $4.826\times10^{-4}$ at $1$~fs to $1.231\times10^{-4}$ at $0.5$~fs and $3.421\times10^{-5}$ at $0.25$~fs. The corresponding P3M and Ewald values are nearly identical, demonstrating that \ac{p3maze} exhibits the same time-step convergence as the reference electrostatic methods.

The FFT-based P3M implementation and the Ewald solver evaluate the reciprocal-space contribution directly, without introducing an iterative-solver tolerance. Consequently, no residual iterative-solver error is present, and the total momentum is conserved close to machine precision in the present simulations. By contrast, real-space \ac{p3maze} is iterative and the accuracy of its solution depends on the prescribed convergence tolerance. As shown below, tightening this tolerance systematically improves the conservation of the total momentum in our \ac{p3maze} simulations, approaching the level observed for the direct reciprocal-space methods.

\begin{figure}[H]
    \centering
    \includegraphics[width=0.82\textwidth]{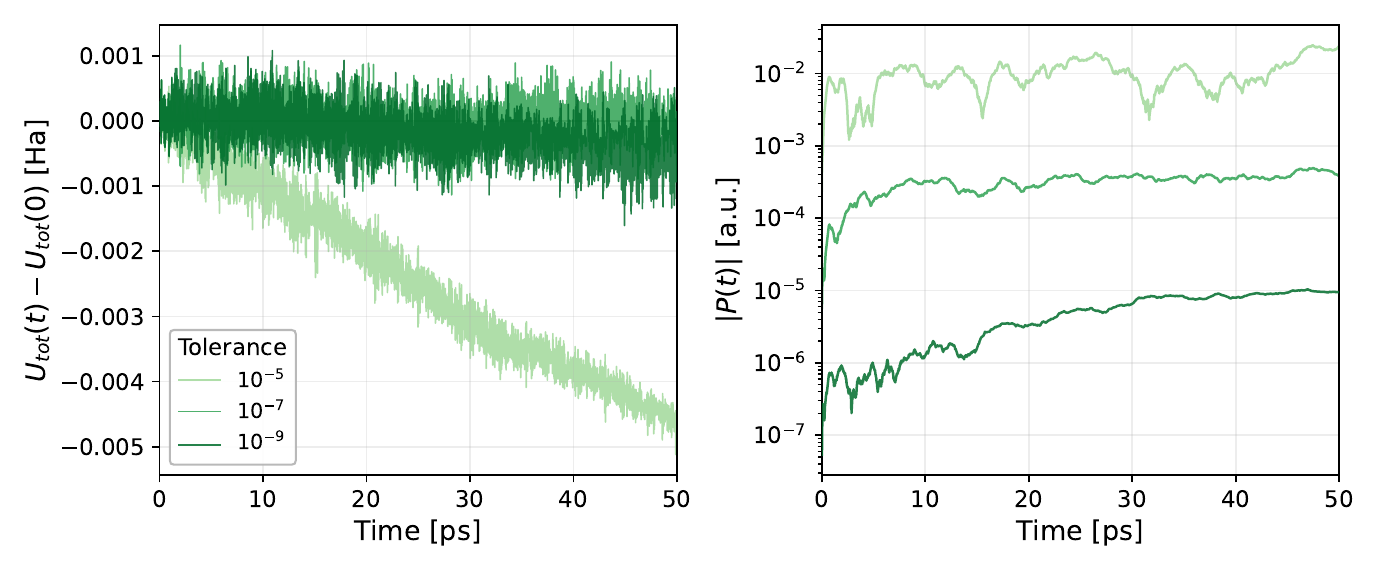}
    \caption{
    Dependence of the \ac{p3maze} NVE stability on the constraint tolerance at fixed time step $\Delta t=0.5$~fs. The left panel shows the total-energy change evaluated with the work-based estimator, while the right panel shows the total-momentum norm. Both energy and momentum conservation improve as the constraint tolerance is tightened.
    }
    \label{fig:water_nve_tolerance}
\end{figure}

We next examined the effect of the \ac{p3maze} constraint tolerance at fixed $\Delta t=0.5$~fs. Figure~\ref{fig:water_nve_tolerance} compares tolerances of $10^{-5}$, $10^{-7}$, and $10^{-9}$. A tolerance of $10^{-5}$ produces a visible systematic energy drift, whereas the trajectories obtained with $10^{-7}$ and $10^{-9}$ reduce or eliminate this drift over the  $50$~ps interval. The relative total-energy fluctuation decreases from $5.317\times10^{-4}$ at $10^{-5}$ to $1.231\times10^{-4}$ at $10^{-7}$. Tightening the tolerance further to $10^{-9}$ gives a comparable fluctuation of $1.398\times10^{-4}$, indicating that further tightening of the constraint tolerance provides no measurable improvement at this integration time step.

Momentum conservation also improves systematically as the tolerance is tightened. This behavior shows that the residual momentum accumulation is associated with the finite accuracy of the constrained mesh solution and can therefore be controlled through the solver tolerance. Unlike other real-space approaches, our formulation does not require an explicit force correction to enforce total-momentum conservation. For example, in the real-space multigrid method of Sagui and Darden~\cite{Sagui:2001}, the force-interpolation procedure does not strictly conserve momentum, and the average net electrostatic force is explicitly subtracted from each particle at every MD step. No analogous force correction was applied in our \ac{p3maze} simulations. The center-of-mass momentum was removed only once, at the beginning of each trajectory, and was not subsequently corrected during the dynamics. The reported momentum therefore reflects its free evolution from an initially zero-momentum state. As discussed in the section on force evaluation, the present formulation preserves total momentum in the converged solution without requiring an explicit correction~\cite{Hockney:1981}.

The relative total- and potential-energy fluctuations are summarized in Table~\ref{tab:water_nve_relative_fluctuations}. We quantify the energy fluctuations through the relative standard deviations $\Delta U_{\rm tot} /|\langle U_{\rm tot}\rangle|$, and $\Delta U /|\langle U_{\rm tot}\rangle|$, where $U_{\rm tot}$ is the total energy, $U$ is the total potential energy, and $\Delta U_{\rm tot}$ and $\Delta U$ denote their sample standard deviations over the complete $50$~ps trajectory. The time-step results show comparable total-energy fluctuations for \ac{p3maze}, P3M, and Ewald. The tolerance sweep shows that a constraint tolerance of $10^{-7}$ is sufficient to make the \ac{p3maze} energy fluctuations comparable to those of the reference methods at the same time step.

\begin{table}[H]
\centering
\caption{
Relative fluctuations of the total energy, $\Delta U_{\rm tot}/|\langle U_{\rm tot}\rangle|$,
and potential energy, $\Delta U/|\langle U_{\rm tot}\rangle|$, for the NVE
simulations of SPC/Fw water. For \ac{p3maze}, the work-based and explicit $\frac{1}{2}q\phi$ energy
definitions are reported separately; for \acs{p3m} and Ewald, $U$ is
the total potential energy. The tolerance sweep is reported only for \ac{p3maze}.}
\label{tab:water_nve_relative_fluctuations}
\footnotesize
\setlength{\tabcolsep}{5pt}
\renewcommand{\arraystretch}{1.12}
\begin{tabular}{lllcc}
\hline\hline
Method
& Value
& Energy definition
& $\Delta U_{\rm tot}/|\langle U_{\rm tot}\rangle|$
& $\Delta U/|\langle U_{\rm tot}\rangle|$ \\
\hline
\multicolumn{5}{l}{\textbf{Time step}} \\
\ac{p3maze} & 1 fs & $E_{\mathrm{work}}$ & $4.826\times10^{-4}$ & $9.312\times10^{-3}$ \\
\ac{p3maze} & 1 fs & $\frac{1}{2}q\phi$ & $8.468\times10^{-4}$ & $9.199\times10^{-3}$ \\
\acs{p3m} & 1 fs & Total & $4.867\times10^{-4}$ & $9.453\times10^{-3}$ \\
Ewald & 1 fs & Total & $4.639\times10^{-4}$ & $9.372\times10^{-3}$ \\
\ac{p3maze} & 0.5 fs & $E_{\mathrm{work}}$ & $1.231\times10^{-4}$ & $9.289\times10^{-3}$ \\
\ac{p3maze} & 0.5 fs & $\frac{1}{2}q\phi$ & $7.823\times10^{-4}$ & $9.162\times10^{-3}$ \\
\acs{p3m} & 0.5 fs & Total & $1.140\times10^{-4}$ & $9.172\times10^{-3}$ \\
Ewald & 0.5 fs & Total & $1.133\times10^{-4}$ & $9.502\times10^{-3}$ \\
\ac{p3maze} & 0.25 fs & $E_{\mathrm{work}}$ & $3.421\times10^{-5}$ & $9.039\times10^{-3}$ \\
\ac{p3maze} & 0.25 fs & $\frac{1}{2}q\phi$ & $9.963\times10^{-4}$ & $8.962\times10^{-3}$ \\
\acs{p3m} & 0.25 fs & Total & $3.402\times10^{-5}$ & $9.404\times10^{-3}$ \\
Ewald & 0.25 fs & Total & $3.476\times10^{-5}$ & $9.256\times10^{-3}$ \\
\hline
\multicolumn{5}{l}{\textbf{Tolerance}} \\
\ac{p3maze} & $10^{-5}$ & $E_{\mathrm{work}}$ & $5.317\times10^{-4}$ & $1.849\times10^{-2}$ \\
\ac{p3maze} & $10^{-5}$ & $\frac{1}{2}q\phi$ & $3.588\times10^{-2}$ & $2.153\times10^{-2}$ \\
\ac{p3maze} & $10^{-7}$ & $E_{\mathrm{work}}$ & $1.231\times10^{-4}$ & $9.289\times10^{-3}$ \\
\ac{p3maze} & $10^{-7}$ & $\frac{1}{2}q\phi$ & $7.823\times10^{-4}$ & $9.162\times10^{-3}$ \\
\ac{p3maze} & $10^{-9}$ & $E_{\mathrm{work}}$ & $1.398\times10^{-4}$ & $9.490\times10^{-3}$ \\
\ac{p3maze} & $10^{-9}$ & $\frac{1}{2}q\phi$ & $7.316\times10^{-4}$ & $9.351\times10^{-3}$ \\
\hline\hline
\end{tabular}
\end{table}
As expected for NVE dynamics, the potential-energy fluctuations remain finite and nearly independent of the integration time step, whereas the total-energy fluctuations decrease systematically with the timestep.

\section{Tuning of the \ac{p3maze} parameters}

The \ac{p3maze} decomposition is controlled by the Gaussian screening width $\sigma$, the real-space cutoff $R_c$, and the number of mesh points per direction $N$. These parameters are coupled: narrower Gaussians require finer grids to resolve the screening charge density, but decay more rapidly in real
space and therefore allow a smaller $R_c$~\cite{Sagui:2001}. The production parameters were selected from explicit convergence tests on the molten-salt system, using as reference a highly converged \ac{pmaze} calculation on a grid with $N=380$ points per direction. For fixed $\sigma=1.39~\text{\AA}$, the relative force error was computed for cutoffs in the range $R_c=3.0$--$8.0~\text{\AA}$ (in steps of $0.5~\text{\AA}$). The error decreased monotonically with $R_c$ and reached a plateau at approximately $R_c=4.5~\text{\AA}$, indicating that the residual error is no longer dominated by truncation of the short-range screened interaction. A convergence study in $N$ at fixed $\sigma$ and $R_c=4.5~\text{\AA}$ then showed that $N=60$ ($h=0.344~\text{\AA}$) yields a relative force error of approximately $2.5\%$ with respect to the reference at moderate computational
cost. These values were adopted for production simulations. The same analysis was performed for the \acs{spc} water model.

\section{\Acs{spc} parameters}
Table adapted from~\cite{wu:2006}.
\begin{table}[H]
\centering
\caption{\ac{spc} model parameters adopted in this work, taken from Ref.~\cite{wu:2006}.}
\label{tab:spcfw_params}
\footnotesize
\setlength{\tabcolsep}{6pt}
\renewcommand{\arraystretch}{1.12}
\begin{tabular}{lcc}
\hline\hline
Parameter & Symbol & Value \\
\hline
Oxygen Lennard-Jones energy  & $\epsilon_{\mathrm{OO}}$  & $0.1554253\ \mathrm{kcal/mol}$ \\
Oxygen Lennard-Jones size    & $\sigma_{\mathrm{OO}}$    & $3.165492\ ~\text{\AA}$ \\
Oxygen charge                & $q_{\mathrm{O}}$          & $-0.82\,e$ \\
Hydrogen charge              & $q_{\mathrm{H}}$          & $+0.41\,e$ \\
Equilibrium O--H bond length & $r_{\mathrm{OH}}^0$       & $1.012\ ~\text{\AA}$ \\
Equilibrium H--O--H angle    & $\theta_{\mathrm{HOH}}^0$ & $113.24^\circ$ \\
Bond force constant          & $k_b$                     & $1059.162~\mathrm{kcal\,mol^{-1}}\,\text{\AA}^{-2}$ \\
Angle force constant         & $k_a$                     & $75.90\ \mathrm{kcal\ mol^{-1}\ rad^{-2}}$ \\
\hline\hline
\end{tabular}
\end{table}

\section{\Acs{spc} rotational relaxation times}
\begin{figure}[H]
    \centering
    \includegraphics[width=0.8\linewidth]{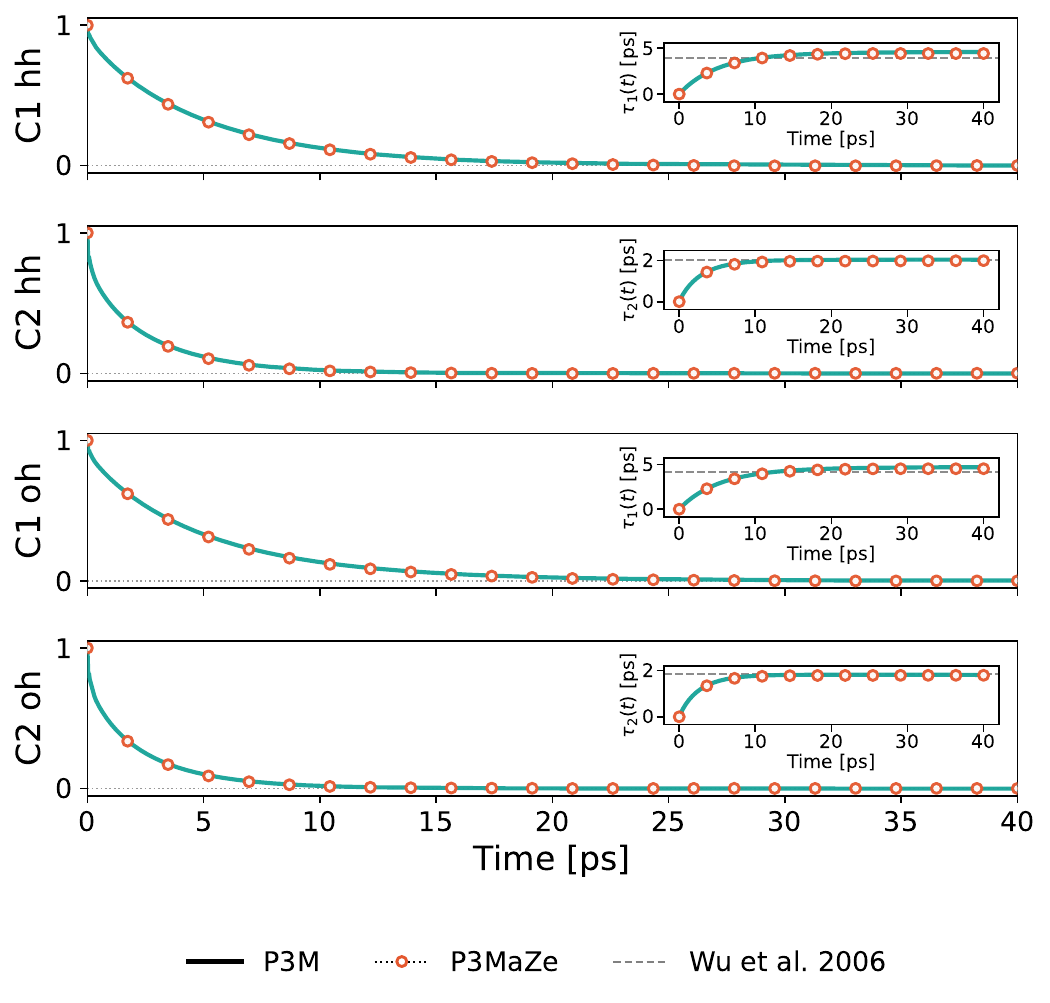}
    \caption{Comparison of rotational relaxation in \ac{spc} water for \ac{p3m} and \ac{p3maze}, considering only the H–H and O–H molecular axes. For each observable, the main panels report the ensemble-averaged first- and second-order rotational correlation functions, while the insets show the corresponding cumulative integral estimates of the relaxation times $\tau_1(t)$ and $\tau_2(t)$. The horizontal dashed lines mark the reference values from Wu \textit{et al.}~\cite{wu:2006}.}
    \label{fig:rotrelax_water_tot}
\end{figure}

\section{Direct comparison of the long-range electrostatic potential and its
translational dependence}
\label{sec:si-potential-translation}

We performed a direct comparison of the long-range electrostatic potentials obtained with our in-house real-space \ac{p3m} implementation and \ac{p3maze}, using the equilibrated 250-ion molten NaCl configuration employed for the molten salt analysis. As shown in the main text, the \ac{p3maze} correction problem and the direct \ac{p3m} Poisson solve are algebraically equivalent representations of the same long-range mesh problem. Here, we provide an additional numerical verification of this equivalence by comparing the resulting electrostatic potentials and their dependence on rigid translations relative to the mesh. The ions were contained in a cubic box of side $L=20.64$~\AA. Both methods used the same $64^3$ mesh, cubic B-spline charge assignment, Gaussian charge smoothing with $\sigma=1.39$~\AA, a real-space cutoff of $4.5$~\AA, and a solver tolerance of $10^{-7}$. The $64^3$ mesh was used specifically for this comparison; the molten NaCl production simulations used the $60^3$ mesh reported in the main text. All calculations were performed with our in-house code to ensure consistent numerical treatment.

No \acl{md} propagation was performed: the particle coordinates were frozen and a fresh solver instance was initialized for every calculation. The complete configuration was rigidly translated relative to the mesh along $x$, $y$, $z$, and the body diagonal. For each direction, 21 equally spaced positions were considered, spanning one mesh spacing in each translated Cartesian component and including the periodic endpoint. After each field update, we stored the vector $\bm{\phi}_{\rm LR}$ of mesh potentials, whose component $\phi_{{\rm LR},n}$ is the potential at grid point $\bm r_n$, with $n=1,\ldots,N_g$ and $N_g=64^3$.
Because the electrostatic potential is defined up to an additive constant, the spatial mean was removed from every field. Translated fields were then registered to the unshifted coordinate system using periodic Fourier interpolation. The registered fields are used throughout the following comparison.

\subsection{Distribution of the \ac{p3m}--\ac{p3maze} potential difference}

At each mesh point, the registered potentials were averaged over 80 fields, comprising 20 samples from each of the four translation directions after excluding the periodic endpoint. We then calculated the vector of potential differences
\begin{equation}
 \Delta\bm{\phi}_{\rm LR}^{\mathrm{methods}}
 =
 \left\langle
 \bm{\phi}_{\rm LR}^{\mathrm{P3M}}
 \right\rangle_{\mathrm{shifts}}
 -
 \left\langle
 \bm{\phi}_{\rm LR}^{\mathrm{P3MaZe}}
 \right\rangle_{\mathrm{shifts}}.
 \label{eq:si-method-delta-phi}
\end{equation}

\begin{figure}[htbp]
  \centering
  \includegraphics[width=0.68\textwidth]{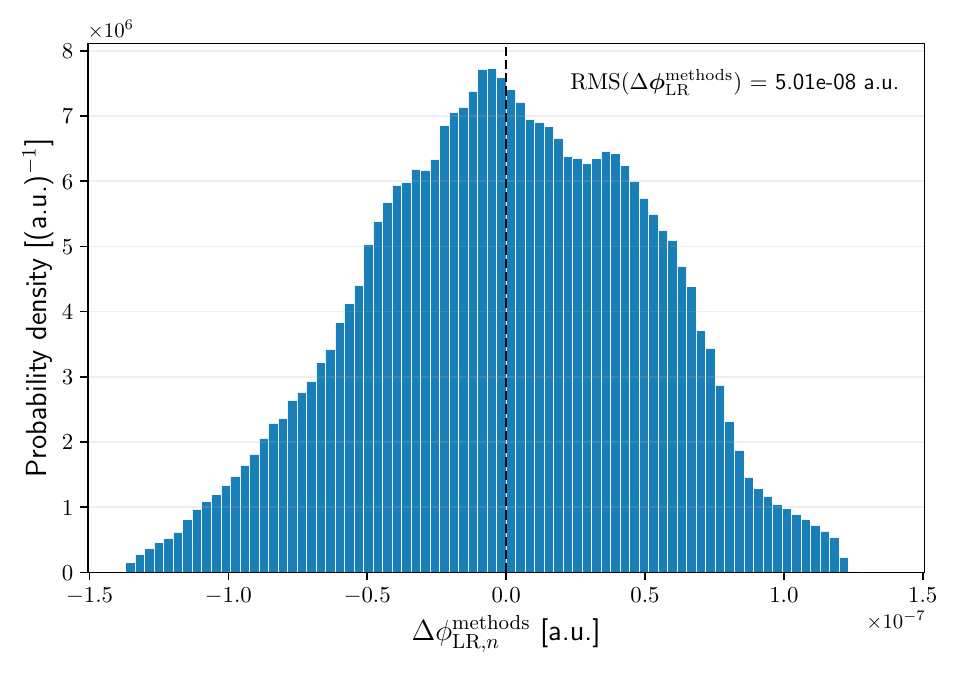}
  \caption{Probability density of the components $\Delta\phi_{{\rm LR},n}^{\mathrm{methods}}$ of the difference between the long-range potentials averaged over translations, evaluated over the $64^3$ mesh points. The vertical dashed line denotes zero. The spatial mean was removed from each potential, and the fields were registered to the unshifted coordinate system before averaging over 80 fields, comprising 20 samples per translation direction after excluding the periodic endpoint.}
  \label{fig:si-potential-histogram}
\end{figure}

Figure~\ref{fig:si-potential-histogram} shows the probability density of the components $\Delta\phi_{{\rm LR},n}^{\mathrm{methods}}$ over the $N_g$ mesh points. The distribution is centred at zero, with a mean of $-2.6\times10^{-20}$~a.u. The RMS difference between the two fields averaged over translations is
\begin{equation}
 \operatorname{RMS}
 \left(\Delta\bm{\phi}_{\rm LR}^{\mathrm{methods}}\right)
 =
 \left[
 \frac{1}{N_g}\sum_{n=1}^{N_g}
 \left(\Delta\phi_{{\rm LR},n}^{\mathrm{methods}}\right)^2
 \right]^{1/2}
 =
 5.0\times10^{-8}\ \mathrm{a.u.},
\end{equation}
and the maximum absolute difference is $1.4\times10^{-7}$~a.u.
The distribution is not required to be Gaussian because its entries are spatially correlated values of a deterministic field rather than independent realizations of a random error. These values quantify the discrepancy between the two numerical methods, rather than an absolute error relative to an analytical reference.

\subsection{Translation of the complete configuration relative to the mesh}

For each method and a rigid translation $\boldsymbol{\Delta}$, we define the registered potential difference at grid point $\bm r_n$ as
\begin{equation}
 \Delta\phi_{{\rm LR},n}(\boldsymbol{\Delta})
 =\phi_{\rm LR}^{(\boldsymbol{\Delta})}(\bm r_n+\boldsymbol{\Delta}) -\phi_{\rm LR}^{(\bm 0)}(\bm r_n),
\end{equation}
where $\phi_{\rm LR}^{(\boldsymbol{\Delta})}(\bm r)$ denotes the
potential reconstructed by periodic Fourier interpolation from the mesh solution for the translated configuration, after removal of its spatial mean. The vector $\Delta\bm{\phi}_{\rm LR}(\boldsymbol{\Delta})$ collects these pointwise differences. We quantify its magnitude through
\begin{equation}
 \operatorname{RMS}
 \left(\Delta\bm{\phi}_{\rm LR}(\boldsymbol{\Delta})\right)
 =
 \left[\frac{1}{N_g}\sum_{n=1}^{N_g} \left(\Delta\phi_{{\rm LR},n}(\boldsymbol{\Delta})\right)^2
 \right]^{1/2}.
 \label{eq:si-rms-translation}
\end{equation}

\begin{figure}[htbp]
  \centering
  \includegraphics[width=0.98\textwidth]{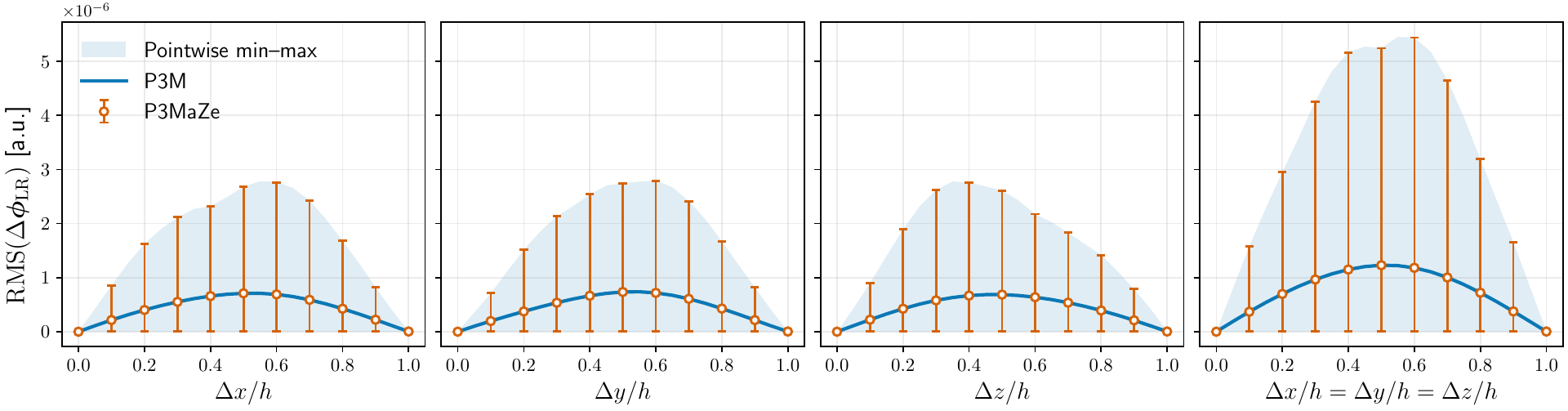}
  \caption{Translational dependence of the long-range electrostatic
  potential for \ac{p3m} and \ac{p3maze}. The ordinate is the RMS potential
  difference defined in Eq.~\eqref{eq:si-rms-translation}. The abscissae
  are $\Delta x/h$, $\Delta y/h$, and $\Delta z/h$ for the Cartesian scans;
  for the body diagonal,
  $\Delta x/h=\Delta y/h=\Delta z/h$. A value of 1 corresponds to a
  translation of one mesh spacing along each indicated component.
  The shaded region and error bars span the range between the minimum
  and maximum of
  $|\Delta\phi_{{\rm LR},n}(\boldsymbol{\Delta})|$ over the mesh points.}
  \label{fig:si-potential-translation}
\end{figure}

Figure~\ref{fig:si-potential-translation} reports the RMS potential difference as a function of the translation components in units of the mesh spacing $h$. For the diagonal scan,
$\Delta x/h=\Delta y/h=\Delta z/h$. The \ac{p3m} and \ac{p3maze} curves are essentially indistinguishable. The RMS variation is zero for the unshifted configuration, is largest near a half-mesh translation, and returns to nearly zero after one complete mesh spacing in each translated component, as expected from mesh periodicity. Its maximum is approximately $7\times10^{-7}$~a.u. for translations along the individual Cartesian directions and $1.23\times10^{-6}$~a.u. for the body diagonal, where all three Cartesian components are varied simultaneously. The shaded region for \ac{p3m} and the error bars for \ac{p3maze} span the range between the minimum and maximum of $|\Delta\phi_{{\rm LR},n}(\boldsymbol{\Delta})|$ over the mesh points. They describe a spatial range and are not statistical confidence intervals from independent trajectories.

The RMS difference between the potentials averaged over translations is smaller than the maximum RMS variation induced by translations relative to the mesh. The agreement between the averaged fields and the coincident translational dependence are consistent with the algebraic equivalence between the \ac{p3maze} correction problem and the corresponding direct Poisson solve established in the main text.

\section{Linear performance fits for Fig.~\ref{main-fig:performance_scaling}}
Table~\ref{tab:performance_fits} reports the coefficients obtained by fitting the untransformed solver and total electrostatic CPU times per step as linear functions of the number of particles, \(t=aN_p+b\). These fits complement the log--log representation shown in Fig.~\ref{main-fig:performance_scaling} of the main text: the slopes in the log--log representation characterize the scaling exponent, whereas the coefficients \(a_{\rm solver}\) and \(a_{\rm elec}\) quantify the increase in solver and total electrostatic cost, respectively, per additional particle. 

\begin{table}[H]
\centering
\caption{
Slopes obtained from linear fits of the untransformed solver time,
$t_{\rm solver}$, and total electrostatic CPU time, $t_{\rm elec}$,
as functions of the number of particles $N_p$. All coefficients are reported in units of $s$. The ratios are normalized by the corresponding slope of \ac{p3maze} with $\bm{y}_0=\bm{y}_0^{(1)}$.}
\label{tab:performance_fits}
\footnotesize
\setlength{\tabcolsep}{5pt}
\renewcommand{\arraystretch}{1.12}
\begin{tabular}{lcccc}
\hline\hline
Method
& $a_{\rm solver}$
& $a_{\rm solver}/a_{\rm solver}^{\rm \ac{p3maze}}$
& $a_{\rm elec}$
& $a_{\rm elec}/a_{\rm elec}^{\rm \ac{p3maze}}$ \\
\hline
\multicolumn{5}{l}{\textbf{NaCl}} \\
\acs{p3maze} - $\bm{y}_0=\bm{y}_0^{(0)}$
& $2.076\times10^{-4}$
& $2.80$
& $2.814\times10^{-4}$
& $1.78$ \\
\ac{p3maze} - $\bm{y}_0=\bm{y}_0^{(1)}$
& $7.423\times10^{-5}$
& $1.00$
& $1.584\times10^{-4}$
& $1.00$ \\
\ac{pmaze}
& $1.383\times10^{-3}$
& $18.63$
& $1.387\times10^{-3}$
& $8.76$ \\
\hline
\multicolumn{5}{l}{\textbf{\ac{spc}}} \\
\acs{p3maze} - $\bm{y}_0=\bm{y}_0^{(0)}$
& $2.627\times10^{-5}$
& $1.04$
& $3.324\times10^{-5}$
& $1.03$ \\
\ac{p3maze} - $\bm{y}_0=\bm{y}_0^{(1)}$
& $2.518\times10^{-5}$
& $1.00$
& $3.216\times10^{-5}$
& $1.00$ \\
\ac{p3maze} - $\bm{y}_0=\bm{y}_0^{(2)}$
& $1.722\times10^{-5}$
& $0.68$
& $2.419\times10^{-5}$
& $0.75$ \\
\ac{p3maze} - $\bm{y}_0=\bm{y}_0^{(3)}$
& $1.514\times10^{-5}$
& $0.60$
& $2.212\times10^{-5}$
& $0.69$ \\
\hline\hline
\end{tabular}
\end{table}
The fitted coefficients \(a_{\rm elec}\) quantify the increase in electrostatic cost per additional particle for each method and system. For \ac{spc} water, the total-cost ratio relative to \ac{p3maze} with \(\bm{y}_0=\bm{y}_0^{(1)}\) decreases from \(1.00\) to \(0.75\) for \(\bm{y}_0^{(2)}\) and \(0.69\) for \(\bm{y}_0^{(3)}\). Thus, the higher-order predictors reduce the total cost per particle, although the incremental improvement becomes smaller with increasing predictor order. This diminishing benefit can be attributed partly to the smoothing overhead, whose contribution increases from approximately \(13\)--\(18\%\) for \(\bm{y}_0^{(1)}\) to \(24\)--\(26\%\) for \(\bm{y}_0^{(3)}\), thereby increasingly limiting the gains obtained by further improving the initial guess.
\bibliography{main}